\documentclass[10pt]{amsart}
\usepackage{amsaddr}

\usepackage{epsfig,amsmath,amsthm,amssymb}
\usepackage{color}

\usepackage{color}
\usepackage{graphicx}

\usepackage{cancel}

\usepackage[colorlinks=true, pdfstartview=FitV, linkcolor=blue, citecolor=blue, urlcolor=blue]{hyperref}
\definecolor{shadecolor}{rgb}{0.95, 0.95, 0.86}

\def\blue#1{\textcolor[rgb]{0,0,1}{#1}}

\begin{document}

\def\ra{\rightarrow}
\newcommand{\al}{\alpha}
\newcommand{\bt}{\beta}
\newcommand{\de}{\delta}
\newcommand{\ga}{\gamma}
\newcommand{\G}{\Gamma}
\renewcommand{\O}{\Omega}
\renewcommand{\k}{\varkappa}
\renewcommand{\d}{\delta}
\newcommand{\D}{\Delta}
\newcommand{\e}{\epsilon}
\renewcommand{\o}{\omega}
\newcommand{\g}{\gamma}
\newcommand{\la}{\lambda}
\newcommand{\si}{\sigma}
\newcommand{\ph}{\varphi}
\renewcommand{\part}{\partial}
\newcommand{\vn}{\varnothing}
\newcommand{\coi}{C_0^{\infty}}
\newcommand{\ci}{C^{\infty}}
\newcommand{\ts}{\text{supp}\,}
\newcommand{\tss}{\text{singsupp}\,}
\newcommand{\br}{{\mathbb R}}
\newcommand{\rn}{\mathbb R^n}
\newcommand{\adx}{\vct\cdot x}
\newcommand{\Sn}{S^{n-1}}
\newcommand{\sg}{\text{sgn}}
\newcommand{\ibr}{\int_{\br}}
\newcommand{\brt}{\br^2}
\newcommand{\irt}{\int_{\brt}}
\newcommand{\ioi}{\int_0^{\infty}}
\newcommand{\iii}{\int_{-\infty}^{\infty}}
\newcommand{\isn}{\int_{\Sn}}
\newcommand{\fle}{f_{\Lambda\e}}
\newcommand{\CS}{{\mathcal S}}
\newcommand{\CD}{{\mathcal D}}
\newcommand{\CH}{{\mathcal H}}
\newcommand{\CB}{\mathcal B}

\newtheorem{theorem}{Theorem}[section]
\newtheorem{example}{Example}[section]
\newtheorem{exercise}{Exercise}[section]
\newtheorem{examps}{Examples}[section]
\newtheorem{lemma}{Lemma}[section]
\newtheorem{remark}{Remark}[section]
\newtheorem{problem}{Problem}[section]
\def\mod{\,\hbox{mod}\,}
\newtheorem{remarks}[remark]{Remarks}
\newtheorem{proposition}{Proposition}[section]
\newtheorem{corollary}{Corollary}[section]
\newtheorem{definition}{Definition}[section]
\def\le{\left}
\def\ri{\right}
\def\ds{\displaystyle}
\def\V {\mathcal V}

\def\un{\underline}
\def\res{\mathop{\mathrm {res}}\limits_}

\def\bt{\begin{theorem}}
\def\et{\end{theorem}}
\def\bc{\begin{corollary}}
\def\ec{\end{corollary}}
\def\ord{\hbox{ ord }}
\def\bx{\begin{example}}
\def\ex{\end{example}}
\def\bxs{\begin{examps}\small}
\def\exs{\end{examps}}
\def\bxr{\begin{exercise}\small}
\def\exr{\end{exercise}}
\def\bl{\begin{lemma}}
\def\el{\end{lemma}}
\def\bd{\begin{definition}}
\def\ed{\end{definition}}
\def\bp{\begin{proposition}}
\def\ep{\end{proposition}}

\def\br{\begin{remark}}
\def\er{\end{remark}}

\def\be{\begin{equation}}
\def\ee{\end{equation}}
\def\ov {\overline}
\def\bes{$$}
\def\ees{$$}
\def\&{\hspace{-15pt}&}
\def\bea{\begin{eqnarray}}
\def\eea{\end{eqnarray}}
\def\beas{\begin{eqnarray*}}
\def\eeas{\end{eqnarray*}}
\def\B{\boldsymbol {\mathfrak B}}
\def\bs{\boldsymbol}
\def \pa{\partial}
\def\C{{\mathbb C}}
\def\L{\mathcal L}
\def\R{{\mathbb R}}
\def\N{{\mathbb N}}
\def\wh{\widehat}
\def\H{{\cal H}}
\def\Z{{\mathbb Z}}
\def\u{\mathfrak u}
\def\a{\alpha}
\def\b{\beta}
\def\g{\gamma}
\def\tgt{\hat{\tilde\gamma}}
\def\gt{\hat\gamma}
\def\K{\mathcal K}
\def\m{\mu}
\def\l{\lambda}
\def\Pot{\mathcal {P}}
\def\1{{\bf 1}}
\def\r{\rho}
\def\c{ {\mathbf c}}
\def\s{ {\sigma}}
\def\t{ {\tau}}
\def\tg{ {\tilde\gamma}}
\def\th{ {\theta}}
\def\w{{\mathbf w}}
\def\x{\xi}
\def\z{\zeta}
\def\ch{\mathbf {C}}
\def\sh{\mathbf {S}}
\def\hf{\frac{1}{2}}

\newcommand{\Iscr}{\mathcal I}
\newcommand{\Bscr}{\mathcal B}
\newcommand{\Lscr}{\mathcal L}
\newcommand{\Wscr}{\mathcal W}
\newcommand{\Mscr}{\mathcal M}
\newcommand{\Nscr}{\mathcal N}
\newcommand{\Rscr}{\mathcal R}
\numberwithin{equation}{section}

\title[Semiclassical NLS: Whitham equations and the Riemann-Hilbert Problem]{Semiclassical limit of the focusing NLS: Whitham equations and the Riemann-Hilbert Problem approach}
\author{Alexander Tovbis}
\address{Department of Mathematics,   University of Central Florida, 4000 Central Florida Blvd. P.O. Box 161364, Orlando, FL 32816-1364}
\email{Alexander.Tovbis@ucf.edu}
\author{Gennady A. El}
\address{ Department of Mathematical Sciences, Loughborough University, 
Loughborough, LE11 3TU, UK}
\email{G.El@lboro.ac.uk}

\maketitle


\begin{abstract} 
The main goal of this paper is to  put together: a) the Whitham theory applicable to slowly modulated
$N$-phase nonlinear wave solutions to the focusing nonlinear Schr\"odinger (fNLS) equation, and b) the Riemann-Hilbert Problem approach to particular solutions  of the  fNLS in the semiclassical (small dispersion) limit that develop slowly modulated $N$-phase nonlinear wave in the process of evolution.  Both approaches have their own merits and limitations. 
Understanding of the interrelations between them could prove beneficial for  a broad range of problems involving the semiclassical fNLS.

\end{abstract}

\maketitle

\section{Introduction}\label{sec-intro}

In this paper we consider the focusing Nonlinear Schr\"odinger (fNLS) equation 
\be  \label{FNLS}
i\varepsilon\partial_t \psi +\hf \varepsilon^2\partial_x^2 \psi + |\psi|^2\psi=0, 
\ee
where $x\in\R$ and $t\ge 0$ are space-time variable and $\varepsilon>0$. In the semiclassical (small dispersion) limit we take
$\varepsilon\ra 0^+$. The fNLS is  a basic model  for self-focusing and
self-modulation, for example, it governs nonlinear  transmission of light in optical fibers; 
it can also be derived as a  modulation equation for  a broad class  of nonlinear systems. 
It was first integrated (with $\varepsilon=1$) by Zakharov and Shabat \cite{ZS},
who produced a Lax pair for it and used the inverse scattering
procedure to describe the time evolution of general decaying potentials ($\lim_{|x|\to \infty}\psi(x,0,\varepsilon)=0$) in terms of the scattering data, that is,
radiation  and solitons. 

Integrability of the fNLS enables the existence of the finite-gap solutions, the non-decaying quasi-periodic potentials with many remarkable properties \cite{belokolos}.
Slow modulations of these potentials are described by the Whitham modulation equations \cite{whitham_book}, a system of quasilinear equations derived by the averaging procedure applied to
fNLS.    The remarkable feature of the fNLS evolution is that the modulated finite-gap potentials arise as asymptotic solutions  in the initial value problems for the fNLS with non-oscillating initial conditions (e.g. Gaussian or sech initial data for (\ref{FNLS})). This has been understood in the framework of the Riemann-Hilbert problem (RHP) approach to the inverse scattering  yielding both the finite-gap potentials and their modulations within the rigorous  asymptotic description of the semiclassical fNLS evolution. However, the connections between the Whitham modulation theory and the RHP approach have not been properly traced.

The main results of the present paper: 

\begin{itemize}

\item We  link the main objects of the Whitham theory (Riemann invariants, their characteristic velocities, fundamental differentials $dp,dq$)
with the corresponding objects coming from  the RHP approach, which are expressed in terms of the $g$-function from the RHP approach and the 
corresponding hyperelliptic Riemann surface $\Rscr$;

\item We derive several equivalent forms of the modulation equations (by which we mean  systems  of transcendental equations
for the branchpoints of $\Rscr$), and use them to show that various forms of the generalized hodograph solutions of the Whitham equations
are satisfied;

\item we use the $g$-function to establish the role of the potential  for the differentials $dp,dq$ in determining the breaking 
(phase transition) curves and double point velocities for the fNLS.

\end{itemize}

The  first (spatial) equation of the Lax pair for the fNLS is known as Zakharov - Shabat (ZS) system
\begin{equation}
\label{ZS}
i\varepsilon \frac{d}{dx} W=
\begin{pmatrix}
z & \psi\\
\bar \psi & -z
\end{pmatrix}W~,
\end{equation}
where $z$ is a spectral parameter  and $W$ is a $2$ by $2$ matrix-function. 

The scattering data, corresponding to the initial data  $\psi(x,0,\varepsilon)$,
consist of the reflection coefficient $r_{0}(z,\varepsilon)$,
as well as
of the points of the discrete spectrum, if any, together with their norming constants. 
The time evolution of the scattering data is simple (\cite{ZS}). 
The corresponding evolution of a given potential is obtained 
through the inverse scattering transformation (IST) of the
evolving scattering data. The inverse scattering transform for the fNLS equation \eqref{FNLS} at
the point $x,t$ can be  
cast as the following matrix Riemann-Hilbert Problem (RHP) in the spectral $z$-plane:
find a $2\times 2$ 
matrix-valued function $m(z)=m(z;x,t,\varepsilon)$, which depends on the asymptotic parameter $\varepsilon$ 
and the external parameters
$x,t$, such that: i) $m(z)$ is analytic in $\C\backslash \G$, 
where the contour $\G=\R$ has the natural orientation; ii)
\be\label{RHPmat}
m_+=m_-
\begin{pmatrix} 
1 + r\bar r  &  \bar r\cr
 r  &1 \cr
\end{pmatrix}
=m_- V~   
\ee
on the contour $\G$, where $r(z,\varepsilon)=r_0(z,\varepsilon) \exp [{2i\over \varepsilon}(2z^2 t +zx)]$
and $m_\pm(z)=\lim_{\d\ra 0}m(z\pm i\d)$ with $\d>0$ and $z\in\R$; iii)
$\lim_{z\ra\infty} m(z)=\1,$
where $\1$ denotes the identity matrix. 
In the presence of solitons 
 the contour $\G$  contains additional small circles around the eigenvalues 
with the corresponding jump-matrices (see, for example, \cite{TVZ1} or \cite{KMM}).

A complementary approach to the analysis of nonlinear dispersive equations is provided by the  Whitham modulation theory which produces a system of quasilinear partial differential equations governing slow modulations of periodic or quasi-periodic  waves.  A prominent area where the Whitham method proved extremely useful is the theory of dispersive shock waves (DSWs) \cite{scholarpedia}, \cite{el_hoefer_2016}

We consider an $N$-phase quasi-periodic solution of the fNLS equation (\ref{FNLS}), 
\begin{equation}\label{ngap}
\psi=\psi_N(\varepsilon^{-1}\eta_1,   \varepsilon^{-1} \eta_2, \dots, \varepsilon^{-1} \eta_N; {\boldsymbol \a}, \bar{\boldsymbol \a}), \quad \hbox{where}  \quad \eta_j= k_j x +\omega_j t + \eta^0_j.
\end{equation}
Here $\boldsymbol \a \equiv \{\a_j\}_{j=0}^N \in \mathbb C^{N}$ and c.c. are the branchpoints of the hyperelliptic Riemann surface of genus $N$ on which (\ref{ngap}) lives \cite{belokolos},
the (normalized by $\varepsilon$) wavenumbers $ k_j$ and the frequencies $ \omega_j$  are defined in terms of  the branchpoints, and  $\eta_j^0$ are arbitrary initial phases. 
Now, if the parameters $\a_j$, $\bar \a_j$  are allowed to vary slowly in space and time they must satisfy the Whitham modulation equations, 
\be \label{whitham-n}
(\a_j)_t  = V_j^{(N)}({\bs{\a}},\bs{\bar\a})(\a_j)_x, \quad (\bar \a_j)_t  = \overline V_j^{(N)}(\bs{\a},\bs{\bar \a})(\bar \a_j)_x, \qquad j=0,\dots, N \, ,
\ee
so that $\a_j$, $\bar \a_j$ are Riemann invariants. 
The characteristic velocities $V_j^{(N)}$, $\overline V_j^{(N)}$ are expressed in terms of the Riemann invariants $\a_j$ through hyperelliptic (for $N\ge 2$) or complete elliptic (for $N=1$)  integrals. For $N=0$ the Whitham system
 (\ref{whitham-n}) has the form
\begin{equation}
\label{whitham0}
(\alpha_0)_t = (\tfrac32 \a_0+ \tfrac12 \bar\a_0)(\alpha_0)_x, \qquad   (\bar \alpha_0)_t = (\tfrac 32 \bar \a_0 + \tfrac 12 \a_0)(\bar\alpha_0)_x\, ,
\end{equation}
and  is equivalent to the  dispersionless limit of  (\ref{FNLS})
\begin{equation}\label{dless}
 \rho_t+(\rho u)_x=0, \quad 
 u_t+uu_x - \rho_x =0 \, ,
\end{equation}
where the Riemann invariants and characteristic speeds in (\ref{whitham0}) are expressed in terms of the hydrodynamic ``density'' $\rho=|\psi|^2 \ge 0$ and ``velocity'' $u=-i \varepsilon^{-1}(\arg \psi)_x$  as
\begin{equation}
\label{alpha}
 \alpha_0 = -(\frac{u}{2} +  i \sqrt{\rho}), \qquad V_0^{(0)}= \tfrac32 \a_0+ \tfrac12 \bar\a_0 =-(u + i \sqrt{\rho}) \, .
 \end{equation}
One can see that the characteristics of (\ref{dless}) are complex unless $\rho=0$ implying nonlinear modulational instability of the NLS equation (\ref{FNLS}) in the long-wave limit, and hence, ill-posedness of the initial-value problem for  (\ref{whitham0})  for all but analytical initial data.  

For $N \ge 1$ the characteristic velocities $V_j^{(N)}$ in (\ref{whitham-n}) are also complex,  however, vanishing of the imaginary parts for some $V_j$'s is possible. This ``partial hyperbolicity'' property makes the Whitham systems for $N \ge 1$ radically different compared to the genus zero case and has major implications in terms of stability of some solutions \cite{box_rogue}.

There are (at least) two ways of looking at the Whitham equations. Originally, they were introduced in \cite{whitham65} as the equations obtained by the averaging of  dispersive conservation laws over the family of periodic or quasi-periodic solutions. Later,  Whitham put his method in the very general variational principle framework \cite{whitham_lagrangian}. Another approach leading to the same set of equations is the multiple-scale (nonlinear WKB) expansions method \cite{luke}. If the original dispersive equation is IST integrable, the associated Whitham system turns out to be also  integrable in the sense which will be explained later on.  

Using the finite-gap theory of the KdV equation  \cite{novikov74}, \cite{lax75}  Flaschka, Forest and McLaughlin \cite{ffm} showed that the endpoints of the spectral bands of quasi-periodic finite-gap potentials of the quantum-mechanical Schr\"odinger operator are Riemann invariants of the Whitham modulation system associated with the KdV equation.  Analogous result for the fNLS equation  was obtained by Forest and Lee \cite{flee86}  and Pavlov \cite{pavlov87}.  

The second context in which the Whitham equations appear is  due to  Lax, Levermore and Venakides  \cite{lax_lev}, \cite{ven85}, \cite{lax_lev_ven_94}  who derived them as the equations governing the zero dispersion limit of the KdV equation.  The dispersion parameter  $\varepsilon$  determines the typical scale of  nonlinear oscillations in the KdV solution so the limit as $\varepsilon \to 0$ exists only in a weak sense. Venakides provided the bridge between the Flaschka-Forest-McLaughlin (wave packet averages) and Lax-Levermore (weak limits) results by developing the higher-order Lax-Levermore theory \cite{ven90}.
Finally, the all-encompassing  approach to the small-dispersion KdV was developed  by Deift, Venakides and Zhou \cite{DVZ}, who introduced  the nonlinear steepest descent method for the   RHP  associated with  the   (semi-classical) inverse scattering problem.

The Whitham equations naturally arise in the RHP construction as the equations governing the evolution of the spectral branch points $\a_j$, $\bar \a_j$. To be precise, the RHP theory yields the {\it hodograph solution} to the Whitham equations as the combination of the {\it moment conditions} and the {\it Boutroux conditions}.  Although the connection of the Whitham equations with the RHP construction of the semi-classical limit  is generally well known  it has not been explored to any depth for the fNLS equation. The apparent reason for such an omission is twofold: (i) the existing rigorous analyses of the IVPs for the  fNLS equation (see \cite{KMM}, \cite{TVZ1} and references therein), performed within the  RHP framework, already contain all the information about slow modulations of the solution so there is no need to recover it with the aid of a more restricted Whitham approach;  (ii) the Whitham equations (\ref{whitham-n}) are elliptic so their application to the problems outside the firmly established facts of the existence and convergence  of the relevant solution was suspect.  Nevertheless, the success of the application of the Whitham theory to dispersively modified  hyperbolic conservation laws, particularly in the DSW  theory for the KdV and defocusing NLS equations (see \cite{el_hoefer_2016} and references therein),  provides a strong incentive for the development of a similar theory for the elliptic, focusing case, which is also supported by the extensive numerical evidence that the key features of the small-dispersion ``hyperbolic'' nonlinear dynamics, such as  the co-existence of smooth and  rapidly oscillating regions and weak convergence,  hold true for at least some cases of the semi-classical fNLS evolution (see, e.g.,  \cite{bronski_kutz}, \cite{ctian},  \cite{box_rogue}).

The modulation theory approach to the description of the small dispersion limit of  integrable ``hyperbolic'' equations, such as KdV or defocusing NLS, involves solution of a nonlinear free boundary problem for the associated Whitham modulation  system via the so-called  {\it matching regularization procedure}. This procedure represents an extension of the original Gurevich-Pitaevskii  method of the modulation description of a DSW in the KdV equation \cite{GP} and prescribes the solution genus increase every time it undergoes a gradient catastrophe. The modulation solutions of different genera are  ``glued'' in a spectial way along the {\it breaking curves} which are free boundaries and whose determination is part of the solution (see e.g. \cite{grava} for the detailed description of the matching regularization procedure for the KdV equation with monotone initial conditions).

The matching regularization procedure can be relatively easily implemented  in the problems involving   the fNLS solutions with $N=0$ and $N=1$. This was done in \cite{el93}, \cite{kamch97}, \cite{box_rogue}
for the fNLS dam-break problem. The modulation solution  of \cite{el93}, \cite{kamch97} was rigorously confirmed in \cite{KenJenk} within the RHP analysis of the semi-classical fNLS with the square barrier initial data.  
The breaks involving  $N \ge 2$ as in the first break for  the sech potential (\cite{KMM}, \cite{TVZ1}), or any higher breaks have not been considered within the Whitham theory with the only exception \cite{box_rogue}, where the modulation solution beyond the second break was used to predict the generation of rogue waves (note, however, that the second break was considered in \cite{LM} in the framework of the RHP approach). 

The above discussion  strongly suggests that it would be highly desirable to develop a method for solving the fNLS-Whitham equations  in  problems involving formation of oscillatory regions characterized by the genus $N \ge 2$. It is also clear that the RHP  analysis, in particular, the nonlinear steepest descent method with the $g$-function mechanism, 
can provide valuable clues to  the structure of the modulation solutions.  Indeed, certain elements of the RHP approach offer an elegant way to
circumvent the matching regularization procedure which could be quite awkward when matching modulation solutions with $N \ge 2$.
Thus, a closer exploration of the interconnections between the RHP theory of the small-dispersion fNLS limit and the counterpart Whitham modulation theory seems a worthwhile task. Indeed, an appropriate combination of the Whitham theory and some key {\it elements} of the RHP  which could be termed a ``formal RHP analysis'', complemented by careful numerical simulations recently proved very effective for solving problems of immediate physical interest \cite{box_rogue}.

The paper is organized as follows. In Sections \ref{sect-Gena},  \ref{sec-RHPapp}  we present the general  Whitham theory
approach and the  RHP approach respectively,  as applied to semiclassical fNLS. These two approaches are compared in the case of genus zero
in Section  \ref{sec-whit0}. In Section \ref{sect-whitn} we derive an explicit expression 
for the $g$-function (in the determinantal form) in higher genera regions
 for the box-type potentials and discuss its connections with the corresponding hyperelliptic Riemann surface $\Rscr$.
Solutions of the Whitham equations in terms of $g$-function are discussed in Section \ref{sec-WEMB}. The results of this section are based on 
Theorem \ref{equiv-mod} about three equivalent forms of modulation equations. 
The results of Sections \ref{sect-whitn}, \ref{sec-WEMB}
in the case of analytic potentials were discussed in Section \ref{sect-anal}. Transitions between the regions of different genera
and  the characteristic velocities along breaking curves are discussed in Section \ref{sect-break}. In particular, we show
that, in the square barrier (``box'') potential case, a pair of collided branchpoints on the breaking curve always {\it has  real  characteristic velocity}.

\section{ Whitham equations and hodograph solution}\label{sect-Gena}

 The Whitham equations for the fNLS \ref{FNLS} can be represented as a single generating conservation equation \cite{flee86}, \cite{pavlov87} 
 \begin{equation}\label{gener_whitham}
 \partial_t dp = \partial_x dq,
 \end{equation}
 where $dp(z , \a, \bar \a)$ and $dq(z, \a, \bar \a)$ are certain meromorphic  differentials of the second kind (the quasimomentum and the quasienergy) on the hyperlliptic Riemann surface $\Rscr$ of genus $N$ defined by the radical
$R(z)=\sqrt{\prod_{j=0}^{N}(z-\a_j)(z-\bar \a_j)}$ where $z$ is the complex spectral parameter. The branch points $\bs{\a}=(\a_0, \dots, \a_{N})$ and  c.c. are the points of simple spectrum of the  periodic Zakharov-Shabat operator (\ref{ZS}).

The quasimomentum and quasienergy differentials $dp$ and $dq$ are uniquely defined by  the following properties \cite{flee86}:

(a) $dp$ and $dq$ have the poles of order two and three respectively at $\infty^\pm$  on $\Rscr$ and no other poles;

(b) the expansions of $dp$ and $dq$ in the local coordinate $z=\z^{-1}$ near $\infty^{\pm}$ are
\begin{equation}\label{poles_pq}
\begin{split}
dp \sim \pm [ - \frac{1}{\z^2} + \hbox{holomorphic part}] \ \hbox{near} \  \infty^{\pm}, \\
dq \sim \pm [ - \frac{2}{\z^3} + \hbox{holomorphic part}] \ \hbox{near} \  \infty^{\pm}
\end{split} 
\end{equation}

(c)  $dp$ and $dq$ satisfy  the   normalization conditions 
 \begin{equation}\label{norm_mero}
\oint \limits_{\gt_j}dp =\oint \limits_{\gt_j} dq =0 \, , \qquad j = 1, \dots, N,
 \end{equation}
 where  $\gt_j$ is a {clockwise} loop around  the branchcut connecting $\a_j$ and $\bar \a_j$ (the $\mathbf{A}$-cycle).
 
 Note that $dp,dq$ are Schwarz symmetrical differentials, so that normalization conditions \eqref{norm_mero} are equivalent to 
 {\em Boutroux normalization conditions} for $dp,dq$: {\em all the cycles of $dp,dq$ on $\Rscr$ are real}.

The integrals over the  $\mathbf{B}$-cycles, canonically conjugated to the $\mathbf{A}$-cycles, give the fundamental wavenumbers $k_j$ and frequencies $\omega_j$:
\begin{equation}\label{kom}
k_j=\oint \limits_{\mathbf{B_j}} dp \, , \quad \omega_j= \oint \limits_{\mathbf{B_j}} dq \, .
\end{equation}
An alternative useful representation for the wave numbers and frequencies in terms of {\it holomorphic} differentials is:
\begin{equation}
\label{kj}
k_j=-4\pi i  \k_{N,j} \, , \qquad \omega_j= -4\pi i  \left[  \tfrac12  \sum \limits_{k=0}^{N} (\a_k + \bar \a_k ) \  \k_{N,j} + \k_{N-1,j} \right], \quad j=1, \dots, N,
\end{equation}
where $\k_{j,k}(\bs{\a}, \bs{\bar \a})$ are the coefficients of the normalized holomorphic differentials found from the system
\begin{equation}
\label{holonorm}
\sum \limits_{i=1}^{N}\k_{m,k} \oint \limits_{\gt_k} \frac{z^{m}}{R (z)} dz = \delta_{mk} \, , \quad m,k = 1, \dots, N.
\end{equation}
Here $\delta_{mk}$ is the Kronecker symbol.

The generating equation (\ref{gener_whitham}) has several fundamental consequences:

(i) By multiplying (\ref{gener_whitham}) by $(z - \a)^{3/2}$ and letting $z \to \a_j$ one obtains the diagonal system (\ref{whitham-n}) with the characteristic speeds $V_j$ given by 
\begin{equation}\label{V_j}
V_j=\frac{dq}{dp}{\bigg |}_{z= \a_j }, \quad \overline V_j=\frac{dq}{dp}{\bigg |}_{z= \bar \a_j }.
\end{equation}

(ii) By expanding (\ref{whitham-n}) near $z = \infty$ we obtain an infinite series of averaged local conservation laws of the form $\partial_t P_j({\bs \a}, {\bs {\bar \a}})= \partial_x Q_j({\bs \a}, {\bs {\bar \a}})$, where $P_j$ are the averaged densities of the NLS conservation laws and $Q_j$ the corresponding averaged fluxes.
Any $2N+2$ of these conservation laws are independent.

(iii) By integrating (\ref{whitham-n}) over each of the $N$ $\mathbf{B}$-cycles we obtain, on using (\ref{kom}), $N$ equations for conservation of waves
\begin{equation}
\label{wncl}
\frac{\partial}{\partial t} k_j (\bs{\a}, \bs{ \bar \a}) =  \frac{\partial}{\partial x} \omega_j (\bs{\a}, \bs{\bar \a}), \qquad  j=1, \dots, N \, .
\end{equation} 

Since  (\ref{wncl}) must be consistent with  (\ref{whitham-n}) we obtain a compact and physically insightful representations for $V_j$'s as nonlinear group velocities is (see \cite{el96}, \cite{ekv2001}, \cite{el_hoefer_2016}).
\begin{equation}
\label{Vjg}
V_j^{(N)} =  \frac{\partial \omega_i}{\partial \a_j} / \frac{\partial k_i}{\partial \a_j} ,    \qquad \hbox{for any} \quad i=1, \dots, N. 
\end{equation}
We note that  equations (\ref{wncl}) represent the consistency conditions in the formal averaging procedure  \cite{whitham_lagrangian, whitham_book}  as well as 
in the  the  WKB-type multiple-scale expansions \cite{luke}, \cite{whitham_book}, \cite{dn89} leading to the same Whitham system (\ref{whitham-n}). 
Within this (general) modulation theory framework the wavenumbera $k_j$ and  the frequences $\omega_j$ in the modulated wave are {\it defined} as  the derivatives of the phase $\eta_j$:
\be \label{kom_local}
k_j=(\eta_j)_x,  \quad \omega_j=(\eta_j)_t, \quad j=1, \dots, N.
\ee
Clearly, the definitions  (\ref{kom_local})  by Clairaut's theorem are consistent with the wave conservation equation
(\ref{wncl}).

The Whitham system (\ref{whitham-n})   can be integrated using the Tsarev generalized hodograph transform \cite{tsarev85}. This method was originally developed for hyperbolic systems of hydrodynamic type but  is equally applicable to elliptic systems. Tsarev's result in the application to our present problem can be formulated as follows. Any  local non-constant solution of the modulation system (\ref{whitham-n}) for a given genus $N$ is given in an implicit form by the system of $N$ algebraic equations with complex coefficients
\begin{equation}
\label{w0}
x+V_j(\bs {\alpha}, \bs{\bar \alpha})t = w_j(\bs{\alpha}, \bs{\bar \alpha}) , \quad  x+\overline V_j(\bs{\alpha}, \bs{\bar \alpha})t = \bar w_j(\bs{\alpha},  \bs{\bar \alpha}),  \qquad j= 1,2, \dots, N,
\end{equation}
where  the characteristic speeds $V_j (\bs{\alpha}, \bs{\bar \alpha}) \equiv V^{(N)}_j (\bs{\alpha}, \bs{\bar \alpha})$ are given by (\ref{Vjg}), (\ref{kj}).  The $2N$ unknown complex functions  $w_j$, $\bar w_j$, $j=1,2, \dots, N$ satisfy the system of  {\it linear} partial differential equations
\be\label{tsarev}
\frac{\partial_{\a_j}w_k}{w_k-w_j}=\frac{\partial_{\a_j}V_k}{V_k-V_j}  \quad \hbox{and \ c.c.};  \quad j, k=1,2, \dots, N, \quad k\neq j \, ,
\ee
where $\partial_{\a_j} \equiv \frac{\partial}{\partial \a_j}$.  System (\ref{tsarev}) is overdetermined but compatible for the NLS case studied here owing to the  integrability of fNLS being preserved under the Whitham averaging  \cite{dn89}.

We now note that, for the solution $q (x,t; \varepsilon)$ of the semi-classical fNLS to have an  asymptotic representation in  the form of the modulated finite-band potential  locally depending on $N$  ``torus'' phases $\varepsilon^{-1}\eta_j=\varepsilon^{-1}(k_jx+\omega_j t + \eta_j^0)$ (see (\ref{ngap})),  and at the same time to satisfy the general kinematic conditions  (\ref{kom_local}) one must require that the ``initial phases'' $\eta_j^0$  are not constants but depend on $x,t$ via the branch points $\bs \a, \bs{\bar \a}$. To this end we introduce  the {\it modulation phase shift  functions}  $\Upsilon_j(\boldsymbol {\a, \bar \a})$ by  $\eta_j^0= -\Upsilon_j(\boldsymbol {\a, \bar \a})$ so that the normalized phases $\eta_j$ assume the form (see \cite{box_rogue}, \cite{el_hoefer_2016})
\be \label{phase_Ups}
\eta_j=k_j x + \omega_j t - \Upsilon_j (\bs \a, \bs{\bar \a}). 
\ee
Then  the definition of the local wavenumber in  (\ref{kom_local})  implies
\be
k_j=\frac{\partial (k_j x+ \omega_j t - \Upsilon_j)}{\partial x}, \quad j=1, 2,
 \ee
which yields
 \be \label{TsaUps}
\frac{\part k_j}{\part \a_m}  x+\frac{\part \omega_j}{\part \a_m} t = \frac{\part \Upsilon_j}{\part \a_m} \, , \quad \hbox{and c.c.} ,\quad j,m=1,2, \dots, N.
 \ee
provided $\partial \a_j / \partial x \ne 0$, $\partial \bar \a_j / \partial x \ne 0$, $j=1,2$.  Here $k_j({\bs \a}, {\bs {\bar \a}})$ and $\omega_j(\bs{\a}, \bs{\bar \a})$ are defined by (\ref{kj}), and $\Upsilon_j$'s are yet to be found.  Note that the second condition (\ref{kom_local}) leads to the same set of equations (\ref{TsaUps}). As we shall see, only half of the equations (\ref{TsaUps}) are independent, so it is sufficient to consider either $j=1$ or $j=2$.  We also note that equations (\ref{TsaUps}) admit a compact and elegant representation in the form of the stationary phase conditions:
\be \label{stat_phase}
\frac{\partial \eta_j }{\partial  \a_m} =0, \qquad \frac{\partial \eta_j }{\partial \bar \a_m} =0, \quad j,m=1,2, \dots, N.
\ee
For given functions $\Upsilon_j(\bs{\a}, \bs{\bar \a})$ equations (\ref{TsaUps}) fully define the modulations $\bs{\a}(x,t)$, $\bs{\bar \a}(x,t)$ (assuming invertibility of (\ref{TsaUps}), which is not guaranteed {\it a priori}). 
Comparing equations (\ref{TsaUps}) with  the hodograph solution (\ref{w0})  and using the representation (\ref{Vjg}) for the characteristic speeds $V_j(\bs \a, \bs{\bar \a})$ in (\ref{w0}) one readily makes the identification
\be \label{wUps}
w_m= \frac{\partial_{\a_m} \Upsilon_j}{\partial_{\a_m} k_j} \quad \hbox{and c.c.}, \quad j,m=1,2, \dots, N \, .
\ee
Now we observe  that formula (\ref{wUps})  must yield the same function $w_m(\bs \a, \bs{\bar \a})$ for all values of $j$.  This is a consequence of the consistency of the genus $N$ Whitham modulation system with $N$ ``extra''  conservation laws (\ref{wncl})  (the same argument  was used to establish the   `nonlinear group velocity' representation  (\ref{Vjg}) for the characteristic speeds of the Whitham modulation system). Thus, it is sufficient to consider any $N$ of the equations (\ref{TsaUps}) for any given $j$.

 Summarizing, the integration of the Whitham equations reduces to the determination of the ``modulation phase shift'' vector function $\bs \Upsilon (\bs \a, \bs{\bar \a})$. As we shall see, this  function naturally arises in the  RHP construction, thus enabling one to circumvent the complicated matching regularization procedure necessary for the determination of the dependence of $\bs \Upsilon (\bs \a, \bs{\bar \a})$ on the fNLS initial conditions within the Whitham modulation theory framework.  Another feature of the RHP analysis enhancing the modulation theory is that it   reveals the precise mechanism  of the genus change across a breaking curve.  We note that within the Whitham modulation theory the genus   change determination is essentially an ``educated guess'' process which must be confirmed by the  construction of the full modulation solution. 

\section {RHP approach to the inverse scattering for the fNLS. The $g$-function.}\label{sec-RHPapp}

It is well known (see, for example \cite{Zh})
that the 
RHP \eqref{RHPmat} has a unique solution $m(z)$ that has
asymptotics $m(z)=\1+{m_1\over z}+O(z^{-2})$ as $z\ra\infty$, 
and that the solution to the NLS \eqref{FNLS} is given by  $\psi(x,t,\varepsilon)=-2(m_1)_{12}$,
where $(m_1)_{12}$ denotes the $(1,2)$ entry of matrix $m_1$. In the case when
 $r_{0}(z,\varepsilon)$ has analytic continuation into the upper halfplane, the RHP for $m(z)$ 
is simplified by factorizing the jump matrix 
\be\label{fact1} 
V=
\begin{pmatrix} 
1 + r\bar r  &  \bar r\cr
 r  &1 \cr
\end{pmatrix}=\begin{pmatrix}
1 & \bar r\\
0 & 1
\end{pmatrix}
\begin{pmatrix}
1 & 0\\
 r & 1
\end{pmatrix}=V_-V_+~~,
\ee 
and, thus,  ``splitting'' the jump condition \eqref{RHPmat} into two jumps: one with triangular jump matrix
$V_+$ along some contour $\G^+$ in the upper halfplane $\bar \C^+$ 
 and the other
 with triangular jump matrix $V_-$ along some contour $\G_-$ in the  lower halfplane $\bar\C^-$
 (here we assume that $\R$ is included in $\bar \C^\pm$).
Contours $\G^\pm$ are deformations of $\R$.
Due to the Schwarz symmetry of the ZS problem, contours $\G^\pm$ can be chosen to be  symmetric 
to each other with respect to the real axis, and we can restrict our attention  only
to the contour $\G^+\subset \bar\C^+$.   
 For simplicity, we assume 
$\G$ to be a simple, smooth (except for a finitely many points) contour without self-intersections.

The central idea of the  Deift-Zhou nonlinear steepest descent method for asymptotic analysis  of RHPs
is proper factorization of a jump matrix accompanied by the proper deformation of the contour $\G$.  
That is why we assumed that $r_0(z;\varepsilon)$ has analytic continuation from $\R$ into $\bar\C^+$.
Another 
essential element in the  nonlinear steepest descent analysis is the concept of $g$-function.
We use the  $g$-function $g(z)=g(z;x,t)$ to define a transformation, that reduces the 
RHP \eqref{RHPmat} to another RHP that,
in the small $\varepsilon$ limit,
has piece-wise constant (in $z$) jump matrices. In some sense, the $g$ function can be compared with the oscillatory phase
function in the WKB method, which reduces a singularly perturbed ODE (like, for example, \eqref{ZS}) to a system that can be 
solved by a power series in the small parameter $\varepsilon$ (or a fractional power of $\varepsilon$).

The $g$ is build for a particular solution of the fNLS \eqref{FNLS}, given by the corresponding scattering
data. Thus, the $g$ function will be defined by  a Schwarz symmetric function $f_0(z)$,
that can be associated with a scaled logarithm of  $r_0(z)$.  
We assume $f_0(z)$ to be 
analytic, or at least, piece-wise analytic, in some Schwarz symmetrical domains that contain $\G^+$ and $\G^-$ respectively. The meaning of piece-wise analyticity will be addressed below. 

Because of Schwarz symmetry,  $f_0(z)$ may have a purely imaginary jump $2i\Im f_0$ on $\R$. Depending 
on whether or not $\Im f_0(z)\equiv 0$ on some interval of $\R$,   $f_0(z)$ may have either one or two
 analytic components in some region containing $\R$.  
In general,  $f_0(z)$  may also depend on $\varepsilon$.

\bxs\label{ex-box_sech}
1) In the case of of the box (barrier) potential $\psi(x,0,\varepsilon)=q\chi_{[-L,L]}$, where $q>0$ is the hight of the box and 
 $\chi_{[-L,L]}$ is the characteristic function of the segment $[-L,L]$ with $L>0$ being the length of the box,
 the (modified) reflection coefficient $r(z,\varepsilon)$
from \eqref{RHPmat} is given by (\cite{KenJenk})
\be \label{ref-box}
r(z;\varepsilon)=\frac{-q\sin\le(\frac{2L\nu(z)}{\varepsilon}\ri)}{\nu(z)\cos\le(\frac{2L\nu(z)}{\varepsilon}\ri)-iz\sin\le(\frac{2L\nu(z)}{\varepsilon}\ri)}e^{\frac{-2tz^2-2xz-2Lz}{\varepsilon}}
=\sum_{k=0}^\infty \r_k(z)e^{i\th_k(z)/\varepsilon},
\ee
 where 
\be\label{the_k}
\theta_k(z)=2tz^2+2(x-L)z+4kL\nu(z),~~~~~~~\r_k(z)=\r_0^{2k-1}(z)(1-\r_0^2(z)), ~~~k=1,\dots,
\ee
with $\nu(z)=\sqrt{z^2+q^2}$ and  $\r_0=\frac{-iq}{\nu(z)+z}$. In this case $f_0(z)$ is piece-wise analytic, taking values  
$f_0(z)=\th_k(z) -2tz^2-2xz$ in different parts of the spectral plane as will be discussed below. Notice that 
$f_0(z)$ does not have a jump along $\R$, so that there is one analytic component of $f_0$ containing $\R$.

2) In the case of a sech potential with phase $\psi(x,0,\varepsilon)= \cosh^{-1-\frac{2i}{\varepsilon}}$, the function
$f_0(z)$, which is the leading order approximation of $-i\varepsilon\ln r_0(z,\varepsilon)$ as $\varepsilon\ra 0$, was calculated (\cite{TVZ1})
as
\be\label{f_0_sech}
f_0(z)=(z-1)\left[i\pi+2\ln(1-z)\right]-2z\ln z, ~~~~~~~~~\Im z \geq 0.
\ee
It follows from \eqref{f_0_sech} that $\Im f_0(z)\neq 0$ as $z\in \R$ except at $z=\pm 1$. Thus, 
the Schwarz symmetrical function $f_0(z)$ has  the jump $2i\Im f_0(z)$ on $\R$, except at $z=\pm 1$.
So,  $f_0(z)$ has two disjoint
 analytic components in a  region surrounding  $\R$. 
\exs

\br\label{rem-discr}
 The Schwarz-symmetric function $f_0(z)$ from \eqref{f_0_sech} in a more general setting can be associated with both the reflection coefficient and/or the 
density function,  defined on the locus of accumulating (in the limit $\varepsilon\ra 0$) discrete eigenvalues.
For example,  the case of $\psi(x,0,\varepsilon)= \cosh^{-1}(x)$, considered  in \cite{KMM}, corresponds to $r(z;\varepsilon)\equiv 0$ if
$\varepsilon=\frac{1}{N}$, $N\in\N$, so that $f_0(z)$ is defined by the density function only.
From the point of view  of this paper, a particular ``source'' of $f_0(z)$ in \eqref{rhpg}-\eqref{f} is  irrelevant.  
\er

\begin{figure}
\begin{center}
\includegraphics[width=0.37\textwidth]{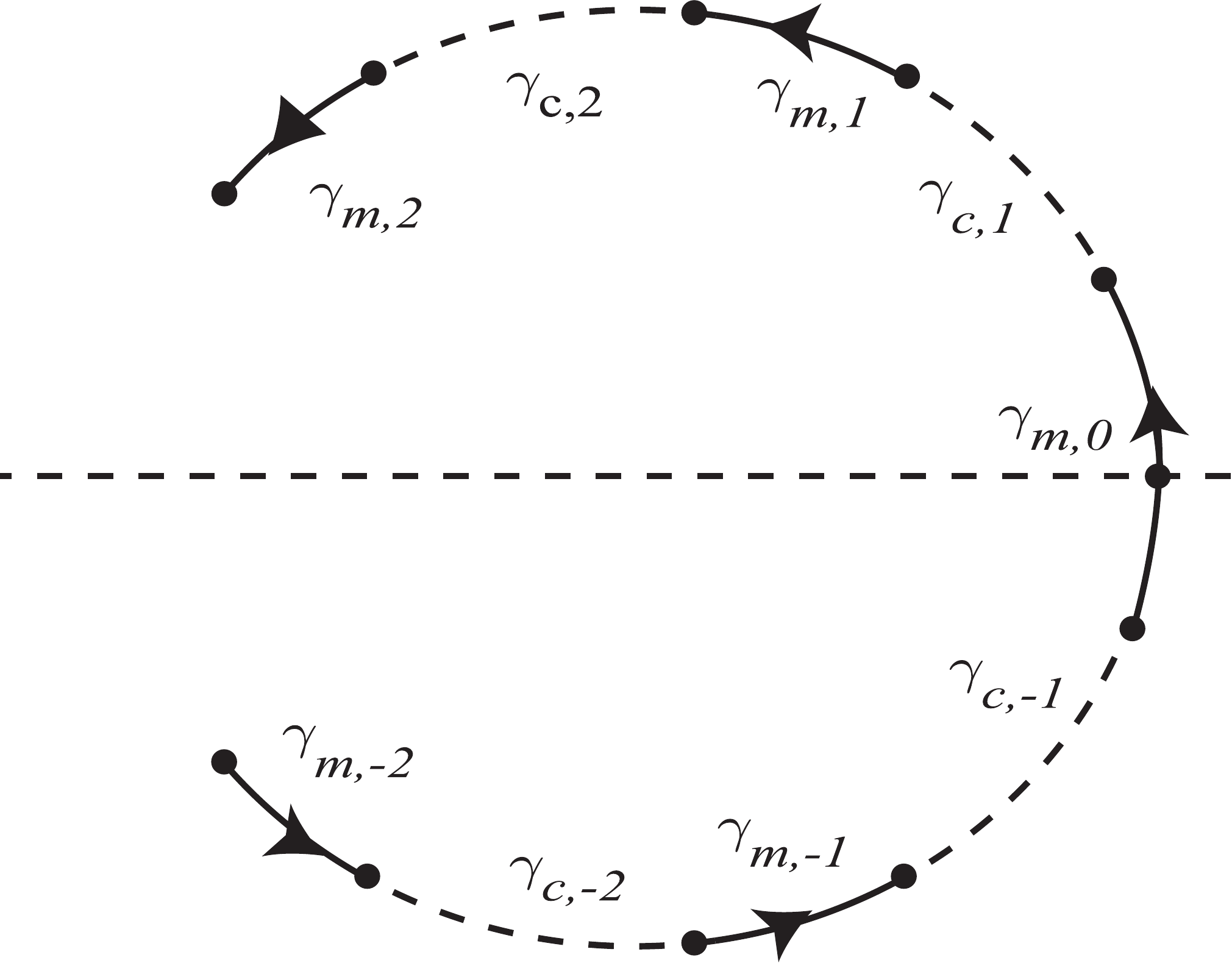}
\includegraphics[width=0.5\textwidth]{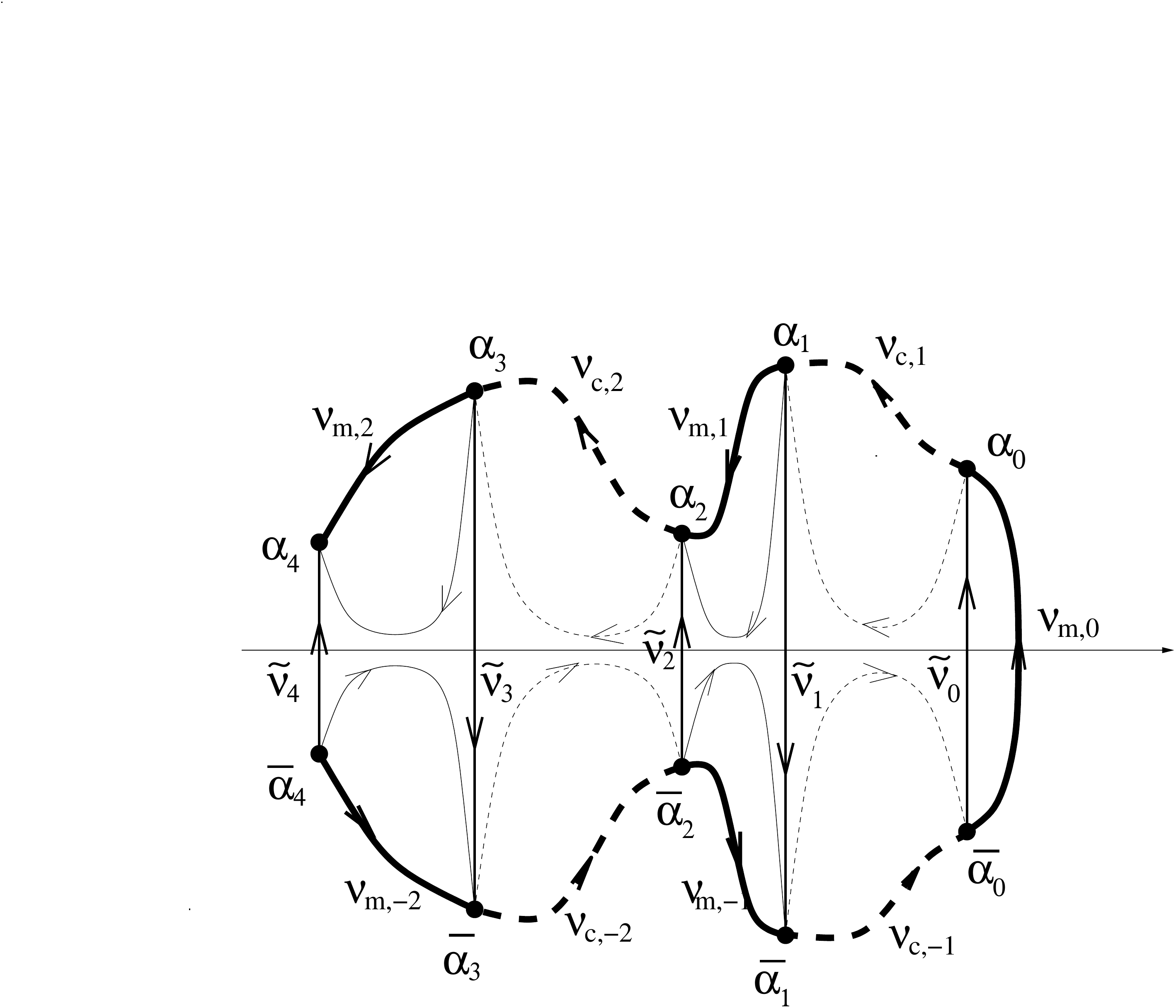}
\caption{Left: Main and complementary arcs with $n=2$  (genus 4) for sech potential, where $f_0(z)$ has a jump across $\R$.
Right: Deformation of main and complementary arcs into vertical branchcuts.}
\label{conturs}
\end{center}
\end{figure}

In order to control the growth of $r(z,\e)$ on the contour $\G$, we split it into a number of arcs of 2 different types:
main arcs (bands) and complementary arcs (gaps). 
Main arcs are always bounded whereas complementary arcs are either unbounded (we do not consider them as they 
do not affect the $g$-function) or bounded arcs that neighbor main arcs at each endpoint.
The number and the positions of the endpoints of these arcs depend
on the initial potential as well as on a particular point $(x,t)$ of the physical (space-time) variables.
The number of bounded complementary arcs (called simply complementary arcs) does not exceed the number $n\in \N$ of main arcs.
Because of Schwarz symmetry, each main arc  either has a complex conjugate (with anti complex conjugate orientation)
or is Schwarz-symmetrical itself, and in this case it crosses the real axis. The same is true for complementary arcs.
We use notations $\g_{m,j}$, $\g_{c,j}$ for the j-th pair of complex conjugate main and complementary arcs respectively
(that also include single self-symmetrical arcs), whereas $\g^\pm_{m,j}$, $\g^\pm_{c,j}$
 denote parts of $\g_{m,j}$, $\g_{c,j}$ in the upper or lower half-planes respectively, see Figure \ref{conturs}, Left.
Here we discuss the general setting of the problem for the $g$-function.

The complex valued Schwarz symmetrical $g$-function is defined as 
satisfying the following  jump and  analyticity conditions:
\begin{align}\label{rhpg}
g_++g_-&=f+W_j ~~\mbox{ on the main arc $\g_{m,j}$, $j=0,\cdots, n$} \cr
g_+ - g_-&=\O_j~~\mbox{ on the complementary arc $\g_{c,j}$, $j=0, 1,\cdots,  n$ }\cr
g(z) &  ~~\mbox{is analytic in ${\bar \C}\setminus \g$, }\cr
\end{align}
where  the function 
\be\label{f}
f(z)=f_0(z)+2zx+2tz^2
\ee
 is a given input to the problem and the  real constants  $W_j$ and $\O_j$ are to be determined (we take $W_0=0$). 
 Note that conditions \eqref{rhpg} form a scalar RHP for the $g$-function $g(z)$.

By the Sokhotski-Plemelj formula,
 \begin{equation}\label{gform1}
g(z)={{R(z)}\over{2\pi i}}
\sum_{j=0}^n\left[ \int_{\g_{m,j}}{{f(\z)+W_j}\over{(\z-z)R_+(\z)}}d\z+
\int_{\g_{c,j}}{{\Omega_j}\over{(\z-z)R(\z)}} d\z\right]~.
\end{equation}
Here the  radical $R(z)=\sqrt{\prod_j(z-\a_j)}$ has branchcuts $\g_{m,j}$, 
where the product is taken over all the endhpoints $\a_j$ of $\g_{m,j}$, $j=0,1,\cdots,n$, 

Let us assume for simplicity that each  $\g_{m,j}$, $j=1,\cdots,n$, consists of two arcs and 
$\g_{m,0}$ is a single arc. Other possible configurations can be considered similarly.
Then the total number of endpoints is $4n+2$ and the hyperelliptic Riemann surface $\Rscr$  of $R(z)$ has the genus 
$N=2n$ and there are no more than $2n$ complementary arcs $\g_{c,j}$, $j=1,\dots,n$. We fix the branch of $R(z)$ by the requirement
\be\label{Rlim}
\lim_{z\ra\infty} \frac{R(z)}{z^{N+1}}=1
\ee
on the main sheet of $\Rscr$.
Due to the analyticity of $f$, the integrals over $\g_{m,j}$, $\g_{c,j}$ in \eqref{gform1}, can be 
deform into the corresponding loop integrals $\gt_{m,j}$, $\gt_{c,j}$ on $\Rscr$ around $\g_{m,j}$, $\g_{c,j}$ respectively.
Thus, 
\be\label{g-hf}
g(z)=O(z-a_k)^\hf
\ee
 around any endpoint $\a_k$ of a main arc. Here and henceforth ee will refer  to the endpoints $\a_j$ as branchpoints (of $\Rscr$).

 The scalar RHP  \eqref{rhpg} for $g(z;x,t)$ is completely  defined if we know the genus $N$ of $\Rscr$,
 the branchpoints $\a_k$ and the real constants $W_j,\O_j$. From the  point of view of solving RHP  \eqref{rhpg},
 the exact location of the main arcs $\g_{m,j}$ is not crucial, as, due to the analyticity of $f(z)$, they can be deformed without changing $g$.
 
 The real constants in the RHP \eqref{rhpg} are determined by the requirement that $g$ is analytic at $z=\infty$ and will be discussed in Section \ref{sect-whitn} below.
 What conditions determine the genus $N$ and the branchpoints? Without going into the details of $g$-function mechanism, 
 we will state that the function
 \be\label{h}
 h(z;x,t)=2g(z;x,t)-f(z;x,t)
 \ee
 must satisfy the following ``sign distribution'' conditions in $\C^+$:
\begin{align}\label{ineq}
\Im h<0& ~~\mbox{ on both sides of each main arc $\g_{m,j}$, $j=0,1,\cdots, n$,} \cr
\Im h>0&~~\mbox{ along each complementary arc $\g_{c,j}$, $j=1,\cdots, n$.}\cr
\end{align}
The corresponding inequalities in the lower half-plane, due to Schwarz symmetry, have the opposite signs.) 
Note that according to \eqref{rhpg}, \eqref{h}, $h_+(z) + h_-(z)=2W_j$ on $\g_{m,j}$, so that
 all the main arcs lie on  zero level curves of $\Im h(z)$.
Conditions \eqref{ineq} can be violated in no more than a finite number of points.
If this is the case, the corresponding value of $(x,t)=(x_b,t_b)$ is called a breaking point.

At any branchpoint, where a main and a complementary arcs meet (movable branchpoint), conditions \eqref{ineq}, combined with \eqref{g-hf}, 
imply that 
\be\label{h-3hf}
\Im h(z;x,t)=O(z-a_k)^{\frac 32}
\ee
Equations \eqref{h-3hf}, are known in the RHP literature as {\em the modulation equations} (although they essentially  are  {\it solutions} of the (differential) Whitham equations), determine the location of all movable (with $x,t$) branchpoints.
Fixed branchpoints (``hard edges'') are known to appear in some non-analytic cases (see, for example, \cite{KenJenk},
for the box initial data). If two or more branchpoints collide at some $\a$, we will have 
\be\label{h-3hf-col}
\Im h(z;x,t)=o(z-a_k)^{\frac 32}
\ee
instead of \eqref{h-3hf}. 
Now, the only remaining question is: how is the genus $N$  defined?

Usually, the genus $N$ is defined for some special values of $x,t$, say, for the initial data $t=0$. 
As we then continuously deform external parameters $x,t$, the branchpoints $\a_j$  move according to
\eqref{h-3hf}, pulling (deforming)  main and complementary arcs of the contour $\g=\g(x,t)$
with them. The genus will be preserved under such deformation until a breaking point $(x_b,t_b)$ is reached.
A {\em regular breaking point is attained by changing the topology of the zero level} of $\Im f(z;x,t)$,  called pinching:
the required inequalities \eqref{ineq} failed in one or several points on the main and/or complementary arcs.
Such points are called double points. Another cause of breaking points is interaction of  the Riemann-Hilbert contour
of $g$ (the collection of the main and complementary arcs) with singularities of $f_0(z)$, or, in the case of piece-wise analytic 
$f_0(z)$, the interaction of the Riemann-Hilbert contour with various elements of $f_0(z)$.

The continuation principle (\cite{TV2}) states that, in the case of a regular breaking point $(x_b,t_b)$,  
we can continue deformation of $x,t$ (with sign conditions satisfied) past $(x_b,t_b)$ with the appropriate
change of the genus $N$.
 
The   scalar RHP \eqref{rhpg} and the modulation equations \eqref{h-3hf} is the starting point of our analysis. 
Various forms of Whitham equations from Section \ref{sect-Gena} and conservation
equations can be derived from \eqref{rhpg} and \eqref{h-3hf}.

\section{The  genus 0 case}\label{sec-whit0}

In the case of $n=0$ there are no real constants in the RHP \eqref{rhpg}. Then the modulation equations \eqref{h-3hf}, defining the branchpoints  of $\Rscr(x,t)$, can be written as
\be\label{modeqint}
\oint_{\gt_m}{{f(\z)}\over{(\z-\a)R(\z)_+}}d\z=0~,
\ee
and its complex conjugate, where $\a=\a_0$, $f$ is given by \eqref{f}
and $\gt_m$ is a
negatively  (clockwise) oriented loop around the main arc (band) that intersects $\R$ only at some point $\m$.
Here $R(z)=\sqrt{(z-\a)(z-\bar\a)}$.

Clearly, \eqref{modeqint} and \eqref{f} yield
\be\label{xalpeq}
2x+2(\Re\a+\a)t=-\frac{1}{2\pi i}\oint_{\gt_m}{{f_0(\z)}\over{(\z-\a)R(\z)_+}}d\z~
\ee
together with its complex conjugate. 
Comparison of \eqref{xalpeq} and  the  hodograph equation \eqref{w0} yields
\be\label{hodo0}
V(\a,\bar\a)=\frac 32(\a+\bar\a), ~~~w(\a,\bar\a)=-\frac{1}{4\pi i}\oint_{\gt_m}{{f_0(\z)}\over{(\z-\a)R(\z)_+}}d\z
\ee
and their complex conjugates, where the velocity  coincides with \eqref{whitham0}.  In the genus zero case
there is only one Tsarev equation (\ref{tsarev})
\be\label{Ts_g=0}
\frac{\part_{\bar\a}w}{w-\bar w}=\frac{\part_{\bar\a}V}{V-\bar V}
\ee
and its complex conjugate, that can be shown true by immediate calculation.
Thus, we proved that $\a,\bar\a$ satisfy the corresponding genus zero Whitham equations \eqref{whitham0}. In fact,
\eqref{modeqint} and its complex conjugate  represent
the hodograph  solution to the Whitham equations \eqref{whitham0}. 

According to \eqref{gform1} and  \eqref{h}, in the genus zero case
\be\label{h0}
h(z)=\frac{R(z)}{2\pi i}\oint_{\gt_m}{{f(\z)}\over{(\z-z)R(\z)_+}}d\z,
\ee
as there are no complementary arcs and the constants $W_0$ can be choosen as zero.
Using
\be
\frac{\part}{\part \a}\frac{R(z)}{R(\z)}=\hf\frac{\z-z}{(z-\a)(\z-\a)}\frac{R(z)}{R(\z)}
\ee
and \eqref{h0}, we obtain
\be\label{heqmod}
\frac{\part}{\part \a}h(z)=\frac{R(z)}{2(z-\a)}\cdot \left. \frac{h(\z)}{R(\z)}\right |_{\z=a}
\ee
and its complex conjugate.
Since, according to \eqref{modeqint}, the last fraction is equal to zero, 
we obtain
\be\label{heqmod=}
\frac{\part}{\part \a}h(z)\equiv 0 \Leftrightarrow  {\rm modulation ~equations} .
\ee

According to \eqref{heqmod=} and \eqref{f}, we have
\be\label{h0-hod}
h(z)=h(z;x,t,\a)= x h_x(z) +th_t(z)  + h_0,~~~{\rm where}~~~h_0=
\frac{R(z)}{2\pi i}\oint_{\gt_m}{{f_0(\z)}\over{(\z-z)R(\z)_+}}d\z
\ee
and 
\be\label{dhdx,dhdt}
h_x(z)=2R(z),~~~~~~~~~~~~~~~h_t(z)=2(z+a)R(z)
\ee
were calculated in \cite{TVZ1}. Here $a=\Re \a$
and $h_x,h_t$ are partial derivatives in the corresponding variables, which, according to \eqref{heqmod=},
coincide with
``full'' derivatives on the solutions of modulation equations $\a=\a(x,t)$.
Obviously,
\be\label{geneq0}
\part_x h_t\equiv  \part_t h_x,
\ee
and one can check directly  that this condition is equivalent to 
 to the modulation equation \eqref{modeqint}. Moreover,  $d h_x$, $d h_t$  are meromorphic differential of the second kind
 on $\Rscr$ with the only poles at $\infty^\pm$ and, according to \eqref{poles_pq},
 \be\label{hpq}
 dp = d h_x,~~~dq=d h_t.
 \ee
 Thus, $p=h_x(z), q=h_t(z)$ are primitive functions for the meromorphic differentials $dp,dq$, and $h(z)$ is a potential
 for $p,q$. {So, we come to the conclusion that the topology of zero level curves of the imaginary part of the potential 
 for $p,q$ that is defined by a particular solution,  determines the change of genus of Whitham equations.}

\section{ Determinantal formula for $g$ when $f_0(z)$ is real analytic on an interval of $\R$} 
\label{sect-whitn}

Consider the case when the (generally piece-wise) analytic function $f_0(z)$ does not have a jump on some interval $I\subset\R$. 
Let us assume for the moment that all the branchpoints $\a_j$ in the upper half-plane have distinct real parts
Then,
deforming the main and complementary arcs, as shown in Figure \ref{conturs}, Right,  we can reduce the RHP \eqref{rhpg} for $g(z)$ to 
the equivalent RHP with jumps along the vertical contours $\tilde\g_j$ with the branchpoints $\bar\a_j,\a_j$, $j=0,\dots, N$:

\begin{align}\label{rhpg-vert}
g_++g_-&=f+\tilde\eta_j ~~\mbox{ on the vertical contour $\tg_{j}$, $j=0,\cdots, N$} \cr
g(z) &  ~~\mbox{is analytic in ${\bar \C}\setminus \cup\tg_j$, }\cr
\end{align}
where  the real constants $\tilde\eta_{2j}$
are expressed as differences of constants $W,\O$ on the neighboring main and complementary arcs. For example, we have 
$\tilde\eta_{2j}=W_j-\O_{j+1}$, $\tilde\eta_{2j+1}=W_{j+1}-\O_{j+1}$ for the configuration, shown on Figure \ref{conturs}, Right.
The $g$ functions, defined by the RHPs \eqref{rhpg} and 
\eqref{rhpg-vert} coincide outside the region, encircled by the ``old'' main and complementary arcs and ``new'' vertical cuts
(called bands), see Figure \ref{conturs}, Right. Inside this region, the ``new'' $g(z)$ coincides with: the ``old''  $g(z)$ up to
appropriate real constant $\O_j$ on the positive (left) side of any two neighboring bands $\tg$;  the values of 
  the ``old''  $g(z)$ from the second sheet of  $\Rscr$ up to
appropriate real constant $W_j$ on the negative (right) side of any two neighboring bands $\tg$
(note the opposite orientation of the neighboring bands). Thus, the modulation equations \eqref{h-3hf} remains valid for the ``new'' $g$,  so that
 either  RHP \eqref{rhpg}   or RHP \eqref{rhpg-vert} can be used to calculate the branchpoints $\a_j$,
the Whitham equations, conservation equations, etc. 
 
Introducing now  $\tilde g=g-\frac{\tilde\eta_0}{2}$, we obtain the RHP
\begin{align}\label{rhpg-vert-fin}
\tilde g_++\tilde g_-&=f+ \eta_j~~\mbox{ on the band $\tg_{j}$, $j=0,\cdots, 2n$} \cr
\tilde g(z) &  ~~\mbox{is analytic in ${\bar \C}\setminus \g$, }\cr
\end{align}
where $\eta_{2j}=  \tilde\eta_{2j}-\tilde\eta_0$. In the configuration of Figure \ref{conturs}, Right,
$\eta_{2j}=W_j-\O_{j+1}+\O_1$, $\eta_{2j+1}=W_{j+1}-\O_{j+1}+\O_1$. 
The constants  $\bs{ \eta}=( \eta_1,\dots,\eta_{2n})$ are exactly the constants from 
 the argument of the multi-phase nonlinear wave solution (given through the Riemann Theta function) of the model problem, see \cite{TVZ1}, Section 8. 
They are also introduced in \eqref{ngap} as fundamental phases.
  With a mild abuse of notation,
we will  use  $g(z)$ instead of $\tilde g(z)$  below and use $\Rscr$ to denote the hyperelliptic Riemann surface 
with branchcuts along $\tg_j$. Here and henceforth we also assume that all the bands are oriented upwards.
This orientation does not change any jump conditions \eqref{rhpg-vert-fin}.

\br\label{rem-bend} 
The vertical branchcuts $\tg_j$ of $\Rscr$ can be bent  to avoid intersection of different $\tg_{j}$, $\tg_{k}$ when $\Re \a_j=\Re \a_k$.
They can also be appropriately bent to intersect $\R$ within the interval $I\subset\R$, where $f_0(z)$ is real analytic.
\er
 
. 
 \bx\label{ex-f_0-box}
 In the case of the box potential, the piece-wise analytic function $f(z)$ is defined by
 \be\label{f_0-box}
 f(z)=2tz^2+2z(x-L)+4kL\nu(z)~~~~{\rm on}~~~~\tg_{2k}\cup\tg_{2k+1},
 \ee
 see \cite{BobAl}, where $\nu(z)=\sqrt{z^2+q^2}$. The number of the bands depends on the particular values of $(x,t)$.
 \ex
 
In the case of genus not exceeding one, the $g$-function, defined by \eqref{rhpg-vert-fin}, coincides with the
 one from \cite{KenJenk} up to   a real constant.   
By Sokhotski-Plemelj formula, we have
\begin{equation}\label{gform2}
2g(z)={{R(z)}\over{2\pi i}}\left[
\sum_{j=1}^N \oint_{\tgt_j}{{\eta_jd\z}\over{(\z-z)R(\z)}} +\oint_{\cup\tgt_k}{{f(\z)d\z}\over{(\z-z)R(\z)}}\right],
\end{equation}
where $R(z)=\prod_{j=0}^{2n}\sqrt{(z-\a_j)(z-\bar\a_j)}$, $\tgt_j$ denotes the negatively (clockwise) oriented loop around 
$\tg_j$ and $f(z)$ is analytically continued 
from  $\tg_j$ to  $\tgt_j$. 
We assume the loops $\tgt_j$ do not intersect each other. With a mild abuse of notation, we will keep using $\gt$ instead of $\tgt$.

The $N$ real constants $\eta_j$, defined by the requirement that $g$ from \eqref{gform2} is analytic at infinity, satisfy
the system of $N$ real linear equations
\begin{equation}\label{momentsNLS}
\frac{1}{2\pi i}\oint_{\cup\gt_j}{\z^k {f(\z)}\over{R(\z)}}d\z
+\frac{1}{2\pi i}\sum_{j=1}^N \oint_{\gt_{j}}{{\eta_j\z^k}\over{R(\z)}}d\z=0
\ \ \ \ \
k=0,1,\cdots, N-1.
\end{equation}
It is well known that the matrix of this system 
\be\label{D-box}
D=\left( \begin{matrix}
  \oint_{\gt_{1}}\frac{ d\z}{R(\z) } &
\cdots &  \oint_{\gt_{1}}\frac{\z^{N-1} d\z}{R(\z) }\cr
\cdots & \cdots & \cdots \cr
  \oint_{\gt_{N}}\frac{ d\z}{R(\z) } &
\cdots &  \oint_{\gt_{N}}\frac{\z^{N-1} d\z}{R(\z) }
\end{matrix}\right)
\ee
is invertible and its inverse matrix
\be\label{D_inv-box}
D^{-1}= \left(\k_{m,j}\right)
\ee
consists of the coefficients of the normalized holomorphic differentials 
\be\label{norm-dif}
\o_j(z)=\frac{\k_{N,j}z^{N-1}+\dots+\k_{1,j} }{R(z)}dz=\frac{p_j(z)dz}{R(z)}, ~~~~j=1,\dots, N
\ee
of the Riemann surface $\Rscr$, defined by $\oint_{\gt_m}\o_k=\d_{mk}$, see \eqref{holonorm}.

Consider  now the determinant
\be\label{KNLS_box}
K(z)= \frac{1}{2\pi i}\times
\left| \begin{matrix}
  \oint_{\gt_{1}}\frac{ d\z}{R(\z) } &
\cdots &  \oint_{\gt_{1}}\frac{\z^{N-1} d\z}{R(\z) } &  \oint_{\gt_{1}}\frac{d\z}{(\z-z)R(\z) }\cr
\cdots & \cdots & \cdots  & \cdots\cr
  \oint_{\gt_{N}}\frac{ d\z}{R(\z) } &
\cdots &  \oint_{\gt_{N}}\frac{\z^{N-1} d\z}{R(\z) }
 &\oint_{\gt_{N}}\frac{d\z}{(\z-z)R(\z) }\cr
\oint_{\cup\gt_j}\frac{f(\z)d\z}{R(\z) } & \cdots
&   \oint_{\cup\gt_j}\frac{\z^{N-1} f(\z)d\z}{R(\z) } &
\oint_{\cup\gt_j}\frac{f(\z)d\z}{(\z-z)R(\z) }\cr
\end{matrix}\right|.
\ee
Multiplying the $j$-th row by $\eta_j$ and adding it to the last row for each $j=1,\dots,N$, we obtain, according to \eqref{gform2} and \eqref{momentsNLS},
\be\label{g-det}
2g(z)=\frac{R(z)}{|D|}K(z),~~~z\in S,~~~~h(z)-\eta_j=\frac{R(z)}{|D|}K(z), ~~~~~~~~z~{\rm inside~ the~ loop~ } \gt_j, 
\ee
where $S$ is a region in $\C$, consisting of  the points that are  outside any loop $\gt_j$, $j=0,\dots,N$.

Note that 
\be\label{h-holom}
K(z)=\frac{h(z)-\eta_j}{R(z)}|D|
\ee
is  analytic inside the loop $\gt_j$.  
Thus, according to \eqref{g-det}, \eqref{KNLS_box}, {\em the modulation equations for all
movable branchpoints $\a_k$}
become
\be\label{mod_K}
K(\a_k)=0.
\ee
In the case of {\em distinct branchpoints $\a_j$}, considered in the paper,  we have $K(z)$ having a simple zero at every movable branchpoint $\a_k$.

Equation \eqref{g-det} allows us to represent $g(z)$ near $z=\infty$ as
\be\label{g-inf}
g(z)=\sum_{j=0}^\infty g_jz^{-j}.
\ee
Indeed, expanding $R(z)=z^{N+1}\sum_{j=0}^\infty \hat R_jz^{-j}$, $\frac{K(z)}{2|D|} =z^{-1} \sum_{j=0}^\infty \hat K_jz^{-j}$,
where
\be\label{g-det-inf}
\hat K_m=\frac{-1}{4\pi i|D|}\left| \begin{matrix}
  \oint_{\gt_{1}}\frac{ d\z}{R(\z) } &
\cdots &  \oint_{\gt_{1}}\frac{\z^{N-1} d\z}{R(\z) } &  \oint_{\gt_{1}}\frac{\z^{N+m} d\z}{R(\z) }\cr
\cdots & \cdots & \cdots  & \cdots\cr
  \oint_{\gt_{N}}\frac{ d\z}{R(\z) } &
\cdots &  \oint_{\gt_{N}}\frac{\z^{N-1} d\z}{R(\z) }
 &\oint_{\gt_{N}}\frac{\z^{N+m} d\z}{R(\z) }\cr
\oint_{\cup\gt_j}\frac{f(\z)d\z}{R(\z) } & \cdots
&   \oint_{\cup\gt_j}\frac{\z^{N-1} f(\z)d\z}{R(\z) } &
\oint_{\cup\gt_j}\frac{\z^{N+m} f(\z)d\z}{R(\z) }\cr
\end{matrix}\right|, 
\ee
we obtain 
\be\label{g-ser}
g(z)=z^N \sum_{j=0}^\infty (\sum_{m=0}^j \hat R_m\hat K_{j-m})z^{-j}.
\ee 
Note that \eqref{g-det-inf}
was obtained by factoring $-z^{-1}$ from the last column of \eqref{KNLS_box}
and expanding  $(1-\frac \z z)^{-1}$.
The condition that $g(z) = O(1)$ at infinity implies that the first $N$ coefficients in \eqref{g-ser} are zeroes and we obtain \eqref{g-inf},
where $g_j=\sum_{m=0}^{j+N} \hat R_m\hat K_{j+N-m}$.

We can also use \eqref{KNLS_box} to calculate the fundamental phases $\eta_j$. Indeed,
taking the proper linear combinations of the first $N$ columns of $K$ in \eqref{KNLS_box}, we obtain
\be\label{KNLS_box-redu}
K(z)= \frac{|D|}{2\pi i}\times
\left| \begin{matrix}
1 &
\cdots &  0&  \oint_{\gt_{1}}\frac{d\z}{(\z-z)R(\z) }\cr
\cdots & \cdots & \cdots  & \cdots\cr
0 &
\cdots & 1
 &\oint_{\gt_{N}}\frac{d\z}{(\z-z)R(\z) }\cr
\oint_{\cup\gt_j}\frac{f(\z)p_1(\z)d\z}{R(\z) } & \cdots
&   \oint_{\cup\gt_j}\frac{ f(\z)p_N(\z)d\z}{R(\z) } &
\oint_{\cup\gt_j}\frac{f(\z)d\z}{(\z-z)R(\z) }\cr
\end{matrix}\right|,
\ee
which implies that
\be\label{const-main}
\eta_m=-\oint_{\cup\gt_j}\frac{f(\z)p_m(\z)d\z}{R(\z) }=4\pi i\le[\b_mt+\k_{N,m}x\ri]-\Upsilon_m,
\ee
where (see (\ref{f}))
\be\label{coeff-const}
\b_m=\k_{N,m}\sum_{k=0}^N\Re\a_k+\k_{N-1,m},~~~
\Upsilon_m=-\oint_{\cup\gt_j}\frac{f_0(\z)p_m(\z)d\z}{R(\z) }.
\ee
Comparing (\ref{const-main}) with (\ref{phase_Ups}) one can see that the phases $\Upsilon_m$ in (\ref{const-main})  coincide with the ``modulation phase shifts'' introduced in Section 
\ref{sect-Gena} in order to provide consistency of the ``torus'' phases $\eta_j$ linearly depending on $x$ and $t$ with the kinematic  definitions (\ref{kom_local}) of the local wavenumbers and frequencies.  These phase shifts can be viewed as the key objects of the modulation analysis as they generate via (\ref{TsaUps}) the generalized hodograph solution to the Whitham equations.     
 The RHP approach offers a direct route to finding  the phases  $\Upsilon_m$ in terms of the scattering data $f_0$ of the initial potential, thus enabling one to circumvent the complicated matching regularization procedure.
In particular, for the box potential we have 
\be\label{Ups-box}
\Upsilon_m=4L\left[i\pi \k_{N,m}+ \sum_{k=0}^N\oint_{\gt_k}\frac{[\frac k2]p_m(\z)\nu(\z)d\z}{R(\z)}\ri].
\ee

To calculate \eqref{const-main}- \eqref{Ups-box} we used expansions
\bea\label{expan}
\frac 1{R(z)}= &\&z^{-N-1}+{\Re \sum \a_j}z^{-N-2}+\dots~~~~\cr
(2tz^2+2xz)p_j(z)= &\&2t\k_{N,j}z^{N+1}+\le[2x\k_{N,j}+2t\k_{N-1,j}\ri]z^N+\dots .
\eea
In the next section we will show that expressions for $\Upsilon_m$ in (\ref{coeff-const}) indeed generate, via (\ref{TsaUps}), the required hodograph solution to the Whitham equations.

\section{Whitham equations for movable branchpoints and their solutions } \label{sec-WEMB}

\bt \label{equiv-mod}
Let $\a$ denote an arbitrary branchpoint $\a_j$,  $j=0,1,\dots, N$, or their complex conjugate,  in the set of $2N+2$ distinct branchpoints. 
Then the following statements are equivalent:
1) $K(\a)=0$; 2) $\frac{\part\vec \eta }{\part\a}=0$; 3) $\frac{\part g(z)}{\part \a}\equiv 0$ for all $z\in\C$,
where $\vec\eta=(\eta_1,\dots,\eta_N)$.
\et

\begin{proof} 
Combining \eqref{gform2}, \eqref{g-det}, \eqref{KNLS_box} with the identity
\be\label{idRR}
 \frac{\part}{\part \a} \le[\frac{R(z)}{(\z-z)R(\z)}\ri]=-\frac{R(z)}{2(z-\a)(\z-\a)R(\z)},
\ee 
we obtain
\be\label{gen-eq}
2\frac{\part g(z)}{\part \a}=-\frac{R(z)K(\a)}{2(z-\a)|D|}+{{R(z)}\over{2\pi i}} 
\sum_{j=1}^N  \frac{\part \eta_j}{\part \a}\oint_{\gt_{j}}{\frac{ d\z}{(\z-z)R(\z)}},
\ee
where $z$ is outside all the loops $\gt_{j}$.

Equation \eqref{g-inf} shows that
\be\label{assg_lam}
\frac{\part g(z)}{\part \a}  ~~~{\rm is ~ analytic ~ at }~ \infty.
\ee
Let  $K(\a)=0$. Then \eqref{assg_lam}
implies the system of  linear equations  
\be\label{syst-deta-dlam}
\sum_{j=1}^N  \frac{\part \eta_j}{\part \a}\oint_{\gt_{j}}{\frac{\z^k d\z}{R(\z)}}
= 0,
~~~~k=0,1\dots,N-1
\ee
for  $\frac{\part \vec\eta}{\part \a}$.
Since the matrix $D$ of this system is invertible, see \eqref{D_inv-box}, and  the right hand side is zero, 
the system  \eqref{syst-deta-dlam} has only zero solution.
 Hence, we proved that 1) implies 2).
Similarly, \eqref{assg_lam} combined with $\frac{\part\vec \eta}{\part\a}=0$ imply 3), that is, 2) implies 3).

Let us now assume 3), that is, $\frac{\part g(z)}{\part \a}\equiv 0$ for all $z\in\C$. Then, differentiating in $\a$ the jump
conditions in \eqref{rhpg-vert-fin} (away from the branchpoints), we obtain 
  $\frac{\part\vec \eta }{\part\a}=0$. Now 1) follows from
\eqref{gen-eq}.

\end{proof}

Here and henceforth we assume that the modulation equations $K(\a_j)=0$ and their complex conjugate
hold for all movable branchpoints $\a_j$.
As an immediate consequence of Theorem \ref{equiv-mod}, we obtain the following corollary.

\bc\label{cor-eta}
For any $j=1,\dots,N$ we now have  \eqref{kj}, namely
\be\label{dg/dxt}
\frac{d}{d x} \eta_j=\frac{\part}{\part x} \eta_j=4\pi i\k_{N,j}=k_j,~~~~~~~\frac{d}{d t}\eta_j=\frac{\part}{\part t} \eta_j=
4\pi i\le(\k_{N,j}\sum_{k=0}^N\Re\a_k+\k_{N-1,j}\ri)=\o_j,
\ee 
where $\frac{d}{d x}$, $\frac{d}{d t}$ denote full derivatives and $k_j,\o_j$ are given by \eqref{ngap}.
\ec

The corollary follows directly from Theorem \ref{equiv-mod} and \eqref{const-main}, \eqref{coeff-const}. One can see that expressions (\ref{dg/dxt}) agree with  the kinematic relations (\ref{kom_local}), which are introduced in the modulation theory as {\it definitions} of the local wavenumbers $k_j$ and  the local frequencies $\omega_j$. 
Moreover, Theorem \ref{equiv-mod} implies the stationary phase conditions \eqref{stat_phase} that can be written 
in the form of  hodograph equations \eqref{TsaUps}.
As earlier, the conservation of waves equations \eqref{wncl}   follows from  Clairaut's theorem.

\bc\label{cor-dg/dxt}
If all the branchpoints $\a_j$,  $j=0,1,\dots, N$,
and their complex conjugates are distinct, then 
\be\label{dg/dxt1}
\frac{d}{d x} g(z;x,t)=\frac{\part}{\part x} g(z;x,t),~~~~~\frac{d}{d t} g(z;x,t)=\frac{\part}{\part t} g(z;x,t),
\ee
where $\frac{d}{d x} $  of $g$ denotes the ``full'' derivative in $x$ that include 
$\sum_j \le(\frac{\part}{\part \a_j}\frac{\part \a_j}{\part x} +\frac{\part}{\part \bar\a_j}\frac{\part \bar\a_j}{\part x}\ri)$.
The same holds for $\frac{d}{d t} $ of $g$. 
\ec

According to \eqref{g-det} and Corollary \ref{cor-dg/dxt}, 
\be\label{dg/dxt-a}
2\frac{d}{d x} g(z;x,t)=\frac{R(z)}{|D|}\frac{\part}{\part x}K(z;x,t),~~~~~2\frac{d}{d t} g(z;x,t)=\frac{R(z)}{|D|}\frac{\part}{\part t}K(z;x,t),
\ee
where $z$ is   outside  all of the loops $\gt_{j}$ and 
\be\label{dh/dxt-a}
\frac{d}{d x} h(z;x,t)-(\eta_j)_x=\frac{R(z)}{|D|}\frac{\part}{\part x}K(z;x,t),~~~~~\frac{d}{d t} h(z;x,t)-(\eta_j)_t=\frac{R(z)}{|D|}\frac{\part}{\part t}K(z;x,t),
\ee
where $z$ is  inside any loop, see  \eqref{h}.
Then, direct calculations show that  
\be\label{dKdxn}
\frac{\part}{\part x} K(z)=2\left| \begin{matrix}
  \oint_{\gt_{1}}\frac{ d\z}{R(\z)} &
\cdots &  \oint_{\gt_{1}}\frac{\z^{N-2} d\z}{R(\z)} &  \oint_{\gt_{1}}\frac{d\z}{(\z-z)R(\z)}\cr
\cdots & \cdots & \cdots  & \cdots\cr
  \oint_{\gt_{N}}\frac{ d\z}{R(\z)} &
\cdots &  \oint_{\gt_{N}}\frac{\z^{N-2} d\z}{R(\z)}
 &\oint_{\gt_{N}}\frac{d\z}{(\z-z)R(\z)}\cr
\end{matrix}\right|+\frac{2z|D|}{R(z)}\chi_S(z)
\ee
and
\be\label{dKdtn}
\begin{split}
\frac{\part}{\part t} K(z)=2\left| \begin{matrix}
  \oint_{\gt_{1}}\frac{ d\z}{R(\z)} &
\cdots &  \oint_{\gt_{1}}\frac{\z^{N-3} d\z}{R(\z)} &  \oint_{\gt_{1}}\frac{d\z}{(\z-z)R(\z)}
&  \oint_{\gt_{1}}\frac{\z^{N-1} d\z}{R(\z)}
\cr 
\cdots& \cdots & \cdots & \cdots  & \cdots\cr
  \oint_{\gt_{N}}\frac{ d\z}{R(\z)} &
\cdots &  \oint_{\gt_{N}}\frac{\z^{N-3} d\z}{R(\z)}
 &\oint_{\gt_{N}}\frac{d\z}{(\z-z)R(\z)}&  \oint_{\gt_{N}}\frac{\z^{N-1} d\z}{R(\z)}\cr
\end{matrix}\right|    \\
+2\sum_{j=0}^{2N}\Re\a_j\frac{\part}{\part x} K(z) + \frac{2z^2|D|}{R(z)}\chi_S(z)~.
\end{split}
\ee
where $ \chi_S(z)$ denotes the characteristic function of the set $S$, which contains all the points of $\bar\C$ outside the loops.
Since we will evaluate $\frac{\part}{\part x} K(z)$, $\frac{\part}{\part t} K(z)$ mostly at the branchpoints, we should disregard
the last terms in \eqref{dKdxn}, \eqref{dKdtn} in the formulae below.

\bl\label{lem-an-hxt}
Functions $\frac{\part}{\part x} h(z;x,t)$, $\frac{\part}{\part t} h(z;x,t)$ are meromorphic on $\Rscr$ with the only poles
at $\infty^\pm$ and attaining the values
\be\label{h(alp)}
h_x(\a_j)=h_x(\bar\a_j)=(\eta_j)_x,~~~~~h_t(\a_j)=h_t(\bar\a_j)=(\eta_j)_t,~~~~j=0,1,\dots,N,
\ee
where $\eta_0=0$.
\el

\begin{proof}


According to \eqref{dg/dxt-a} and \eqref{dKdxn}, \eqref{dKdtn}, functions $\frac{\part}{\part x} g(z;x,t)$, $\frac{\part}{\part t} g(z;x,t)$
are analytic on $S$, where analyticity at $z=\infty$ can be shown by differentiating \eqref{g-inf}. 
Let us collapse the loops $\gt_j$
 onto the bands $\tg_j$, $j=0,\dots,N$, that are the branchcuts of $\Rscr$. Then  $\frac{\part}{\part x} g(z;x,t)$, $\frac{\part}{\part t} g(z;x,t)$ 
 will be analytic on $S=\bar\C\setminus\cup_0^\infty\tg_j$.
 
 Let us fix some $z\in\tg_j$ and denote by $g_\pm(z),h_\pm(z)$ the limiting values of $g,h$ as we collapse $\gt_j$
 onto  $\tg_j$.  Then,  according to \eqref{dg/dxt-a}-\eqref{dKdtn},
 \be\label{jumps-dghxt}
 \begin{split}
 &2  (g_x)_+ - [(h_x)_+ -(\eta_j)_x]= 2z +(\eta_j)_x,~~~~
 ~~~~{\rm on }~~\gt_j,
 ~~~{\rm to  ~ the~ left ~ of}~~\tg_j \\
&  2  (g_x)_- - [(h_x)_- -(\eta_j)_x]= 2z +(\eta_j)_x,~~~~ 
 ~~~~{\rm on }~~\gt_j 
 ~~~{\rm to  ~ the~ right ~ of}~~\tg_j,\\
 &(h_x)_+ -(\eta_j)_x + (h_x)_- -(\eta_j)_x =0,~~~   
 ~~~~{\rm on }~~\tg_j, ~~~~
 \end{split}
\ee
so, adding these jumps, we obtain 
\be\label{jumpgx}
(  g_x)_++(  g_x)_-=2z+ (\eta_j)_x~~\mbox{ on the band $\tg_{j}$, $j=0,\cdots, N$}.
\ee
Similarly, 
 \be\label{jumpgx}
 (  g_t)_++(  g_t)_- =2z^2+ (\eta_j)_t~~\mbox{ on the band $\tg_{j}$, $j=0,\cdots, N$}.
 \ee
Note also that, 
 according to \eqref{dg/dxt-a}-\eqref{dKdtn}, the limiting values $(  g_x)_\pm, (  g_t)_\pm$
on the branchcuts   $\tg_j$ are continuous and bounded functions. So, 
$g_x,g_t$ satisfy  the RHPs similar to
\eqref{rhpg-vert-fin}, and, correspondingly, can be written as
\be\label{sol-rhpgxt}
\frac{\part g(z)}{\part x}=z+
{{R(z)}\over{4\pi i}} 
\sum_{j=1}^N  \oint_{\gt_{j}}{\frac{(\eta_j)_x d\z}{(\z-z)R_+(\z)}},~~~~~~
\frac{\part g(z)}{\part t}=z^2+
{{R(z)}\over{4\pi i}} 
\sum_{j=1}^N \oint_{\gt_{j}}{\frac{(\eta_j)_t d\z}{(\z-z)R_+(\z)}}.
\ee
Thus,  
\be\label{hxt-gxt}
\frac{\part h(z)}{\part x}=\frac{\part g(z)}{\part x}-2z, ~~~
\frac{\part h(z)}{\part t}=\frac{\part g(z)}{\part t}-2z^2
\ee
 are meromorphic on $\Rscr$ with the only poles at $\infty^\pm$.
Equations \eqref{h(alp)} follow from \eqref{dh/dxt-a}.
\end{proof}

This lemma could also be proven using the fact that 
the RHP \eqref{rhpg-vert-fin} for $g(z)$ commutes with $\frac{d}{d x} $, $\frac{d}{d t} $, because  
the boundary values $g_\pm$, as well as $(g_x)_\pm, (g_t)_\pm$ belong to $L^2_{loc}$ along the jump contours $\cup_{j=0}^N \tg_j$.

In the particular  case of $N=2$, we have
\be\label{dKdx}
\frac{\part}{\part x} K(z)=
-2\left|\begin{matrix}\oint_{\gt_1}\frac{ d\z}{(\z-z)R(\z)} &\oint_{\gt_1}\frac{ d\z}{R(\z)} \cr
\oint_{\gt_2}\frac{ d\z}{(\z-z)R(\z)} & \oint_{\gt_2}\frac{ d\z}{R(\z)}\end{matrix}\right|
\ee
and
\be\label{dKdt}
\frac{\part}{\part t} K(z)=
2\left|\begin{matrix}\oint_{\gt_1}\frac{ d\z}{(\z-z)R(\z)}
&\oint_{\gt_1}\frac{\z d\z}{R(\z)} \cr
\oint_{\gt_2}\frac{ d\z}{(\z-z)R(\z)} & \oint_{\gt_2}\frac{\z d\z}{R(\z)}\end{matrix}\right|
+2\sum_{j=0}^2 \Re\a_j\frac{\part}{\part x} K(z)
\ee

\bl\label{lem-dKda} 
Let $\a$ denote any movable $\a_j$,   $j=0,1,\dots,N$, or its complex conjugate, and the modulation equations \eqref{mod_K} hold. Then
\be\label{dKdagen}
\frac{\part K(z)}{\part \a}= \le[\frac{\part \ln |D|}{\part \a}+\frac{1}{2(z-\a)}\ri]K(z).
\ee
In particular, if $\b$ denote another movable branchpoint, then
\be\label{dKda}
\frac{\part}{\part \a}K(\b)=0~~~~{\rm if}~~~~\a\neq \b~~~{\rm and}~~ 
\frac{\part}{\part \a}K(\a)=\hf K'(z)|_{z=\a}.
\ee
\el
\begin{proof}
 Formula \eqref{dKdagen} a direct consequence of Theorem \ref{equiv-mod} and \eqref{g-det}, whereas \eqref{dKda}
follows from  \eqref{KNLS_box}, \eqref{mod_K} and analyticity of $K(z)$ at $z=\a_j$. 
\end{proof}

As a consequence of Lemma \ref{lem-dKda} and the modulation equations \eqref{mod_K}, we obtain
\be\label{NLSdK/daa/xt}
\frac{\part K(\a_j)}{\part \a_j}(\a_j)_x=-\frac{\part}{\part x} K(\a_j),~~~~~~~~~~~
\frac{\part K(\a_j)}{\part \a_j}(\a_j)_t=-\frac{\part}{\part t} K(\a_j)~,~~~~~
\ee
for any movable branchpoint $\a_j$ and its complex conjugate.
Thus, we obtain the corresponding Whitham partial differential equations 
\be\label{whitham_g}
(\a_j)_t=V_j({\bs{\a}},\bs{\bar\a})(\a_j)_x,~~{\rm where}~~V_j({\bs{\a}},\bs{\bar\a})=\frac{\frac{\part}{\part t} K(\a_j)}{\frac{\part}{\part x} K(\a_j)}~~
\ee
as it was stated in \eqref{whitham-n}, Section \ref{sec-intro}.

We also obtain the ordinary differential equations 
\be
\part_t (\a_j)=-\frac{2\part_tK(\a_j)}{K'(\a_j)},~~~~~~~\part_x (\a_j)=-\frac{2\part_xK(\a_j)}{K'(\a_j)},
\ee
for (time) trajectories and (space) isochrones of any movable branchpoint. Note that, according to Lemma \ref{lem-dKda},  $(\a_j)_x, (\a_j)_t$
are bounded provided that all the branchpoints are distinct.


Using the fact that $f(z)$ in the last row of the determinant $K$ in \eqref{KNLS_box} is linear in $x,t$, we can rewrite the modulation
equations \eqref{mod_K} for movable  singularities $\a_j$ and their complex conjugates  as
\be\label{hod-1}
{K_x(\a_j)}x+{K_t(\a_j)}t+{K_0(\a_j)}=0,~~~~
\ee
where  $K_0(z)$ is obtained from $K(z)$ by replacing $f(\z)$ with $f_0(\z)$. Equation \eqref{hod-1} is another form of the 
 hodograph equations (see \eqref{w0})
\be\label{hodon}
x+V_jt=w_j,~~~  ~~~{\rm where}  ~~~V_j=   \frac{K_t(\a_j)}{K_x(\a_j)} ~~~{\rm and} ~~w_j = -\frac{K_0(\a_j)}{K_x(\a_j)}.
\ee
Using \eqref{KNLS_box-redu}, we obtain
\bea\label{K_xt}
K_x(z)&\&=-2|D|\sum_{j=1}^N\k_{N,j}\oint_{\gt_j}\frac{d\z}{(\z-z)R(\z)}d\z, \cr
K_t(z)&\&=-2|D|\sum_{j=1}^N\le(\k_{N,j}\Re\sum_{k=1}^N\a_k+\k_{N-1,j}\ri)\oint_{\gt_j}\frac{d\z}{(\z-z)R(\z)}d\z.
\eea
Then, according to  \eqref{hodon}, we obtain the following {\it  new  expressions for the characteristic velocities}
\be\label{velos}
V_j=\frac{(\a_j)_t}{(\a_j)_x}=\frac{K_t(\a_j)}{K_x(\a_j)}= \Re\sum_{k=1}^N\a_k+
\dfrac{\sum_{k=1}^N\k_{N-1,k}\oint_{\gt_k}\frac{d\z}{(\z-\a_j)R(\z)}d\z}{\sum_{k=1}^N\k_{N,k}\oint_{\gt_k}\frac{d\z}{(\z-\a_j)R(\z)}d\z}
\ee
 in terms of the meromorphic differentials on $\Rscr$.  Tsarev equations \eqref{tsarev} for $w_j $ follow immediately from \eqref{hodon}.

\bl\label{lem-pq}
If
\be\label{pq}
p=-\hf h_x= z-g_x(z),~~~q=-\hf h_t=z^2-g_t(z)
\ee
then
differentials $dp,dq$ are fundamental meromorphic differentials, uniqely determined by the 
conditions a)-c) at the beginning of Section \ref{sect-Gena},
see \eqref{poles_pq}, \eqref{norm_mero}. 
\el

\begin{proof}
According to Lemma \ref{lem-an-hxt}, $p,q$ are meromorphic on the hyperelliptic surface $\Rscr$, with the only  poles at
$\infty^\pm$.
Then 
\be\label{dpq}
dp=-\hf h_{xz}dz = (1-g_{xz}(z))dz,~~~dq=-\hf h_{tz}dz=(2z-g_{tz}(z))dz
\ee
are meromorphic differentials of the second kind with the only poles at $\infty^\pm$ satisfying  \eqref{poles_pq}.

To prove the normalization \eqref{norm_mero} of $dp,dq$, we notice that $\tg_j$, $j=1,\dots,N$ are the ${\mathbf{A}}$-cycles of $\Rscr$ and
 \be\label{norm_dp}
 \oint_{\tg_j}dp =-\int_{\bar\a_j}^{\a_j}h_{xz}dz=h_x(\bar\a_j)-h_x(\a_j)=0,
 \ee
 the latter follows from \eqref{h(alp)}.
 Similar argument based on  \eqref{dh/dxt-a}, \eqref{dKdtn} works for $dq$.
 \end{proof}
 
 \bc\label{cor-gencons}
 The generating conservation equation  $\partial_t dp = \partial_x dq$, see \eqref{gener_whitham}, is obviously true,
 as now it  reduces to 
 $h_{xzt}=h_{tzx}$  (or $g_{xzt}=g_{tzx}$).
\ec

Since the cycle $\mathbf{B_j}$ is a path, connecting $\a_j$ with $\a_0$ and returning back on the second sheet of $\Rscr$,
we obtain \eqref{kom} by
\be\label{B-comp}
\begin{split}
 \oint_{\mathbf{B_j}}dp=-\int_{\a_j}^{\a_0}h_{xz}dz=h_x(\a_j)-h_x(\a_0)=(\eta_j)_x=k_j, \\
  \oint_{\mathbf{B_j}}dq=-\int_{\a_j}^{\a_0}h_{tz}dz=h_t(\a_j)-h_t(\a_0)=(\eta_j)_t=\o_j,
\end{split}
\ee
where we used \eqref{h(alp)}.

Finally, to prove \eqref{V_j}, we notice that at any movable branchpoint $\a_j$ (and its c.c.)
\be\label{hzxt}
h_{xz}(z)=\frac{(R^2(z))'}{2|D|R(z)}K_x(\a_j)+O(z-\a_j)^\hf,~~~~~h_{tz}(z)=\frac{(R^2(z))'}{2|D|R(z)}K_t(\a_j)+O(z-\a_j)^\hf,
\ee
as $z\ra \a_j$. Then, according to \eqref{dpq}, \eqref{hzxt} and \eqref{whitham_g},
\be\label{dqdp}
\le. \frac{dq}{dp}\ri|_{\a_j}=\le. \frac{h_{tz}}{h_{xz}}\ri|_{\a_j}=\frac{K_t(\a_j)}{K_x(\a_j)}=V_j.
\ee
So, we proved  \eqref{V_j}.

\section{ Determinantal formula for $g$ when $f_0(z)$ has a jump on $\R$} \label{sect-anal}

In the case when n $f_0(z)$ has a jump on $\R$ we have to work with the RHP \eqref{rhpg} for the $g$-function.
Similarly to \eqref{momentsNLS},
the real  constants  $ W_j,\O_j\in\R$, $j=1,2,\cdots,n$ in
 \eqref{gform1}  are defined by 
\begin{equation}\label{moments}
\sum_{j=1}^n \frac{W_j}{2\pi i}\oint_{\gt_{m,j}}\frac{\z^k d\z}{R(\z)}
+\sum_{j=1}^n  \frac{\O_j}{2\pi i}\oint_{\gt_{c,j}}{{\z^kd\z}{R(\z)}}=\frac 1{2\pi i}\sum_{j=1}^n \oint_{\gt_{m,j}}\frac{\z^k f(\z)d\z}{R(\z)}
 \ \ \ \ \
k=0,1,\cdots,2n-1.
\end{equation}
This is a  system of $N=2n$ real linear equations for $2n$ real unknowns  $ W_j,\O_j$ that can be written as 
\be\label{mom-vect}
(\vec W,\vec\O)D=\vec f,
\ee
where $\vec W,\vec\O\in\R^n$ are row vectors with components $W_j,\O_j$ respectively,  $\vec f\in\R^{2n}$ denotes the row vector
of the right hand sides of \eqref{moments} multiplied by $2\pi i$, and       
\be\label{DNLS}
D=\left(\begin{matrix} \oint_{\gt_{m,1}}\frac{ d\z}{R(\z)} &
\cdots & \oint_{\gt_{m,1}}\frac{\z^{2n-1} d\z}{R(\z)} \cr
\cdots & \cdots & \cdots \cr
 \oint_{\gt_{m,n}}\frac{ d\z}{R(\z)} &
\cdots & \oint_{\gt_{m,n}}\frac{\z^{2n-1} d\z}{R(\z)}
\cr
\oint_{\gt_{c,1}}\frac{ d\z}{R(\z)} &
\cdots &\ \oint_{\gt_{c,1}}\frac{\z^{2n-1} d\z}{R(\z)}
\cr
\cdots & \cdots & \cdots \cr
 \oint_{\gt_{c,n}}\frac{ d\z}{R(\z)} &
\cdots & \oint_{\gt_{c,n}}\frac{\z^{N-1} d\z}{R(\z)}
\cr
\end{matrix}\right)~
\ee
\medskip
It was shown in \cite{TV1} that if all the branchpoints are distinct then  $|D|\ne 0$.

Let us assume, for simplicity, that the function  $f_0(z)$ is analytic in some region containing all the $\g^+_{m,j}$
Introducing the determinant $K(z)=K(z;x,t)$ by
\be\label{KNLS}
K(z)= \frac{1}{2\pi i}\times
\left| \begin{matrix}
  \oint_{\gt_{m,1}}\frac{ d\z}{R(\z)} &
\cdots &  \oint_{\gt_{m,1}}\frac{\z^{2n-1} d\z}{R(\z)} &  \oint_{\gt_{m,1}}\frac{d\z}{(\z-z)R(\z)}\cr
\cdots & \cdots & \cdots  & \cdots\cr
  \oint_{\gt_{m,n}}\frac{ d\z}{R(\z)} &
\cdots &  \oint_{\gt_{m,n}}\frac{\z^{2n-1} d\z}{R(\z)}
 &\oint_{\gt_{m,n}}\frac{d\z}{(\z-z)R(\z)}\cr
  \oint_{\gt_{c,1}}\frac{ d\z}{R(\z)} &
\cdots &  \oint_{\gt_{c,1}}\frac{\z^{2n-1} d\z}{R(\z)}
 &\oint_{\gt_{c,1}}\frac{d\z}{(\z-z)R(\z)}\cr
\cdots & \cdots & \cdots & \cdots \cr
  \oint_{\gt_{c,n}}\frac{ d\z}{R(\z)} &
\cdots &  \oint_{\gt_{c,n}}\frac{\z^{2n-1} d\z}{R(\z)}
&\oint_{\gt_{c,n}}\frac{d\z}{(\z-z)R(\z)}\cr
\oint_{\gt}\frac{f(\z)d\z}{R(\z)} & \cdots
&   \oint_{\gt}\frac{\z^{2n-1} f(\z)d\z}{R(\z)} &
\oint_{\gt}\frac{f(\z)d\z}{(\z-z)R(\z)}\cr
\end{matrix}\right|,
\ee
we observe that, according to \eqref{mom-vect}, \eqref{gform1},
\be\label{gn}
2g(z)=\frac{R(z)}{|D|}K(z),~~~~z~{\rm outside} ~ \gt,  ~~~~h(z)-\eta_j=\frac{R(z)}{|D|}K(z), ~~~~~~~~z~{\rm inside~ } \gt {\rm ~but  ~ outside}~\gt_j 
\ee
for all $j=1,\dots,N$. here
 the negatively oriented contour $\gt$ is going around all the all main arcs
(it is pinched to $\g_{m,0}$ at $\mu$, where $\{\mu\}= \g_{m,0} \cap \R$).

Combining this considerations with \eqref{modeqint}, we obtain a new form of modulation equations
\be\label{modeq-n}
K(\a_{j})=0,~~~~~~~j=0,1,\cdots,4n+1~.
\ee
We can now state  Theorem \eqref{equiv-mod} for the $g$-function, defined by the RHP \eqref{rhpg}.

\bt \label{equiv-mod-f}
Let $\a$ denote an arbitrary branchpoint $\a_j$,  $j=0,1,\dots, 4n+1$,  in the set of $4n+2$ distinct branchpoints. 
Then the following statements are equivalent:
1) $K(\a)=0$; 2) $\frac{\part(\vec W,\vec\O)^t }{\part\a}=0$; 3) $\frac{\part g(z)}{\part \a}\equiv 0$ for all $z\in\C$.
\et

The proof of the theorem is almost identical to that of Theorem \eqref{equiv-mod}. As an immediate consequence 
of Theorem \eqref{equiv-mod-f}, we obtain
\be\label{dh/dxt-j}
\frac{d}{d x} h(z;x,t)=\frac{R(z)}{|D|}\frac{\part}{\part x}K(z;x,t),~~~~~\frac{d}{d t} h(z;x,t)=\frac{R(z)}{|D|}\frac{\part}{\part t}K(z;x,t),
\ee
where $z$ is inside $\gt$ but  outside all the loops $\gt_{m,j}, \gt_{c,j}$.
Without any lost of generality, we can take limit when contour $\gt$ in \eqref{KNLS} becomes infinitely large.
Then, direct calculations show that  
\be\label{dKdxn-j}
\frac{\part}{\part x} K(z)=\left| \begin{matrix}
  \oint_{\gt_{m,1}}\frac{ d\z}{R(\z)} &
\cdots &  \oint_{\gt_{m,1}}\frac{\z^{2n-2} d\z}{R(\z)} &  \oint_{\gt_{m,1}}\frac{d\z}{(\z-z)R(\z)}\cr
\cdots & \cdots & \cdots  & \cdots\cr
  \oint_{\gt_{m,n}}\frac{ d\z}{R(\z)} &
\cdots &  \oint_{\gt_{m,n}}\frac{\z^{2n-2} d\z}{R(\z)}
 &\oint_{\gt_{m,n}}\frac{d\z}{(\z-z)R(\z)}\cr
  \oint_{\gt_{c,1}}\frac{ d\z}{R(\z)} &
\cdots &  \oint_{\gt_{c,1}}\frac{\z^{2n-2} d\z}{R(\z)}
 &\oint_{\gt_{c,1}}\frac{d\z}{(\z-z)R(\z)}\cr
\cdots & \cdots & \cdots & \cdots \cr
  \oint_{\gt_{c,n}}\frac{ d\z}{R(\z)} &
\cdots &  \oint_{\gt_{c,n}}\frac{\z^{2n-2} d\z}{R(\z)}
&\oint_{\gt_{c,n}}\frac{d\z}{(\z-z)R(\z)}\cr
\end{matrix}\right|
\ee
and
\be\label{dKdtn-j}
\frac{\part}{\part t} K(z)=-2\left| \begin{matrix}
  \oint_{\gt_{m,1}}\frac{ d\z}{R(\z)} &
\cdots &  \oint_{\gt_{m,1}}\frac{\z^{2n-3} d\z}{R(\z)} &  \oint_{\gt_{m,1}}\frac{d\z}{(\z-z)R(\z)}
&  \oint_{\gt_{m,1}}\frac{\z^{2n-1} d\z}{R(\z)}
\cr
\cdots& \cdots & \cdots & \cdots  & \cdots\cr
  \oint_{\gt_{m,n}}\frac{ d\z}{R(\z)} &
\cdots &  \oint_{\gt_{m,n}}\frac{\z^{2n-3} d\z}{R(\z)}
 &\oint_{\gt_{m,n}}\frac{d\z}{(\z-z)R(\z)}&  \oint_{\gt_{m,n}}\frac{\z^{2n-1} d\z}{R(\z)}\cr
  \oint_{\gt_{c,1}}\frac{ d\z}{R(\z)} &
\cdots &  \oint_{\gt_{c,1}}\frac{\z^{2n-3} d\z}{R(\z)}
 &\oint_{\gt_{c,1}}\frac{d\z}{(\z-z)R(\z)}&  \oint_{\gt_{c,1}}\frac{\z^{2n-1} d\z}{R(\z)}\cr
\cdots& \cdots & \cdots & \cdots & \cdots \cr
  \oint_{\gt_{c,n}}\frac{ d\z}{R(\z)} &
\cdots &  \oint_{\gt_{c,n}}\frac{\z^{2n-3} d\z}{R(\z)}
&\oint_{\gt_{c,n}}\frac{d\z}{(\z-z)R(\z)}&  \oint_{\gt_{c,n}}\frac{\z^{2n-1} d\z}{R(\z)}\cr
\end{matrix}\right|
+\sum_{j=0}^{4n+1}\a_j\frac{\part}{\part x} K(z)~.
\ee
Collapsing the loops $\gt_{m,j},\gt_{c,j}$ back to the main and complementary arcs $\g_{m,j},\g_{c,j}$ respectively, we
find that $h_x(z)$ satisfy the RHP with jumps and asymptotics
\begin{align}\label{rhph_x}
(h_x)_++(h_x)_-&=2W_j ~~\mbox{ on the main arc $\g_{m,j}$, $j=0,\cdots, n$} \cr
(h_x)_+-(h_x)_-&=2\O_j~~\mbox{ on the complementary arc $\g_{c,j}$, $j=0, 1,\cdots,  n$ }\cr
h_x=-2z +O(1) &  ~~\mbox{as} z\ra \infty\cr.
\end{align}
The similar RHP is satisfied by $h_t(z)$.
It also follows from \eqref{dh/dxt-j}, \eqref{dKdxn-j}, \eqref{dKdtn-j} that $h_x,h_t$ are bounded at the branchpoints.

We can now deform contours  $\g_{m,j},\g_{c,j}$ in the same way as we did in the beginning of Section \ref{sect-whitn},
to obtain jumps of $h_x,h_t$ on $N+1$ vertical bands $\tg_j,$ $j=0,\dots,N$, where $N=2n$.
These jumps $2(\eta_k)_x, 2(\eta_k)_t$ are given by
$(\eta_{2j})_x=(W_j)_x-(\O_{j+1})_x+(\O_1)_x$, $(\eta_{2j+1})_x=(W_{j+1})_x-(\O_{j+1})_x+(\O_1)_x$
and similar expressions  for $2(\eta_k)_t$. Solutions  $h_x(z),h_t(z)$ to these deformed RHPs
are meromorphic functions on the corresponding hyperelliptic surface $\Rscr$ and their differentials, $dp=-\hf h_{xz}dz$,
$dq=-\hf h_{tz}dz$, satisfy conditions a)-c) from Section \ref{sect-Gena}.
Thus, the results of Section \ref{sect-whitn} are also valid in our case.

\section{Phase transitions and characteristic velocities along breaking curves} \label{sect-break}
As we move in the  $(x,t)$ (physical) plane along some (smooth) curve $\varpi$, the (movable) branchpoints $\a_j=\a_j(x,t)$ move in the $z$ (spectral) 
plane according
to the modulation (Whitham) equations. As they move, the function $h(z;x,t)=h^N(z;x,t)$, where $N$ denotes the genus, 
changes according to \eqref{g-det} (or \eqref{gn})  and, at some point
$(x_b,t_b)\in\varpi$, called a breaking point, one (or more) of the 
inequalities \eqref{ineq} can fail at a some point $z_0$ (or at several points). 
Note that if $z_0\not\in\R$, then  \eqref{ineq} should also fail  at $\bar z_0$
due to Schwarz symmetry of $h^N$. Assuming that $h^N(z)$ is analytic at $z_0$, we readily obtain that
\be\label{br-cond}
h^N_z(z_0)=0~~\qquad ~~{\rm and}~~\qquad ~\Im h^N(z_0)=0
\ee
form a system of 3 real equations for 4 real variables ($x,t, z_0$) that determines the breaking curve.
The point $z_0$ is called a {\em double point}.
If $z_0\in \g_{m,j}$, then two   new complementary arcs open at $z_0$ and at $\bar z_0$ as we cross the breaking curve and the genus of $\Rscr=\Rscr(x,t)$
increases by two. (Generically, it will increase only by one if $z_0\in\R$.) Similarly, if  $z_0\in \g_{c,j}$, two new main arcs open at $z_0$ and at $\bar z_0$.
Moving along $\varpi$ through $(x_b,t_b)$ in the opposite direction, we would observe two branchpoints collide at the double point $z_0$ and then disappear.

The function $h^N(z)=h^N(z;x,t)$ is the potential function for the meromorphic on $\Rscr$ functions $p^N=-\hf h^N_x$ and $q^N=-\hf h^N_t$, whose differentials
$dp^N,dq^N$ determine the system of Whitham  equations. This potential function contains the information about the particular fNLS solution.
Changes of the dimension of this system (in the genus of $\Rscr$) are caused by changes in the
topology of zero level curves of $\Im h^N(z;x,t)=0$.

\begin{figure}
\begin{center}
\includegraphics[width=0.28\textwidth]{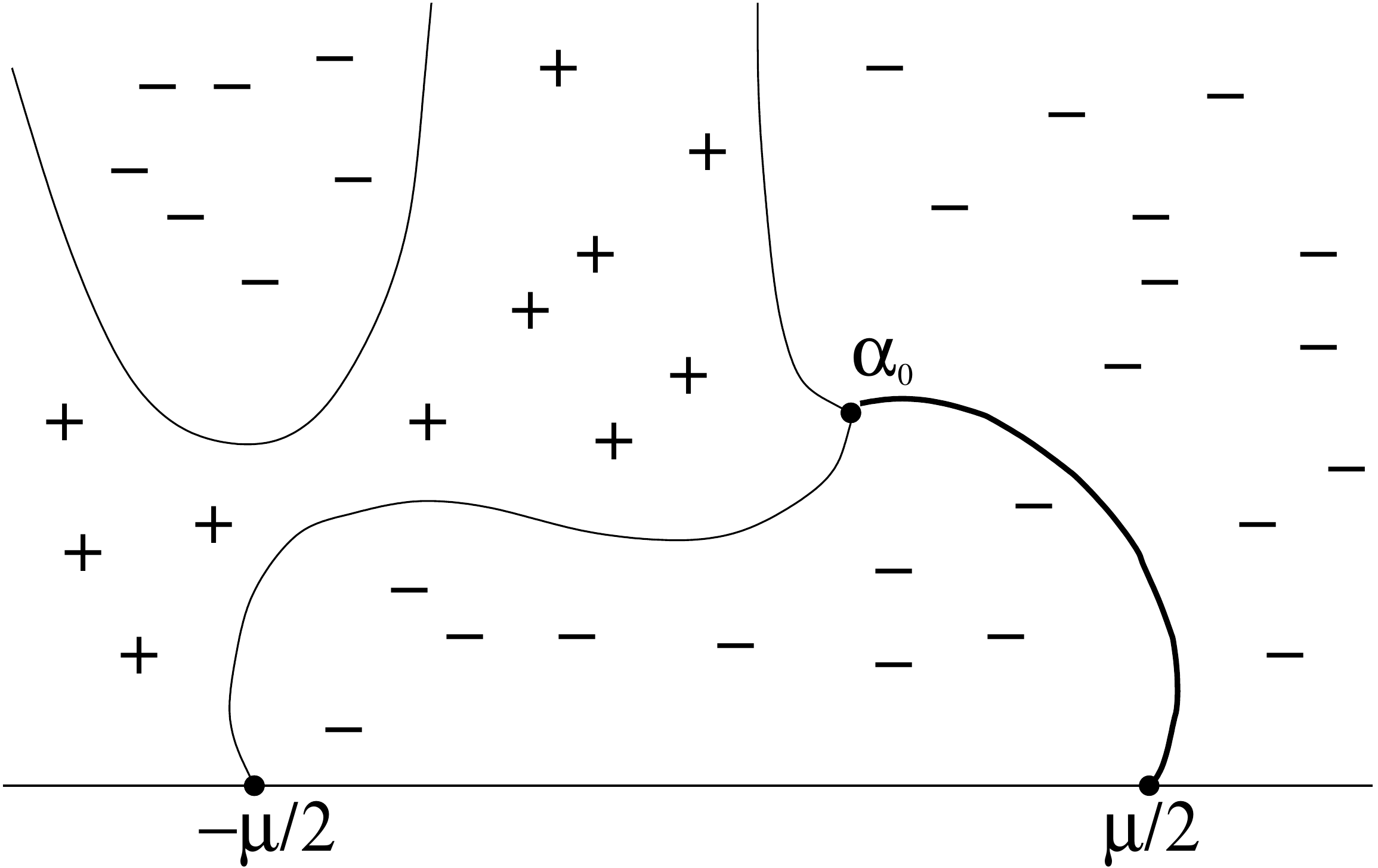}\quad
\includegraphics[width=0.28\textwidth]{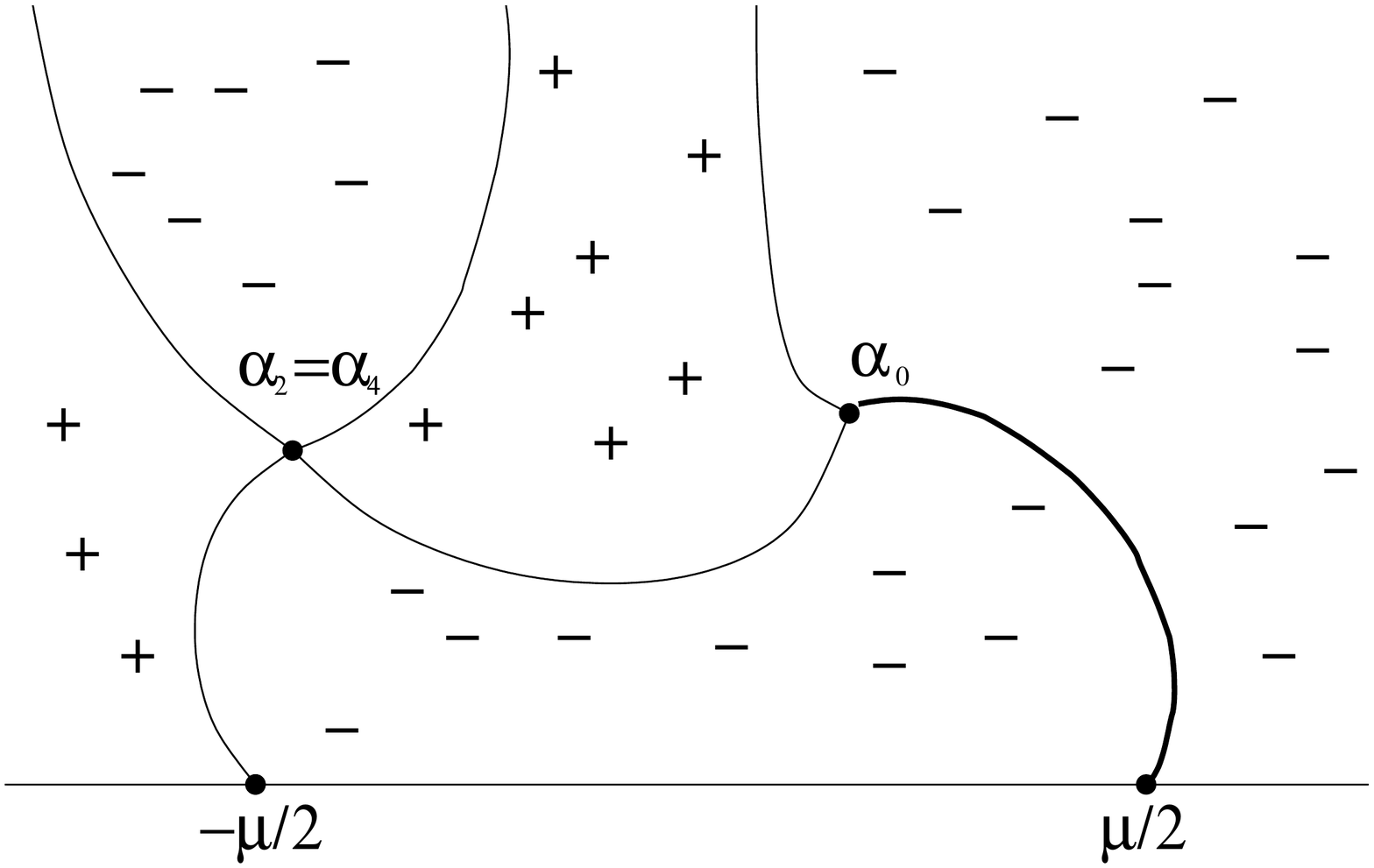}\quad
\includegraphics[width=0.28\textwidth]{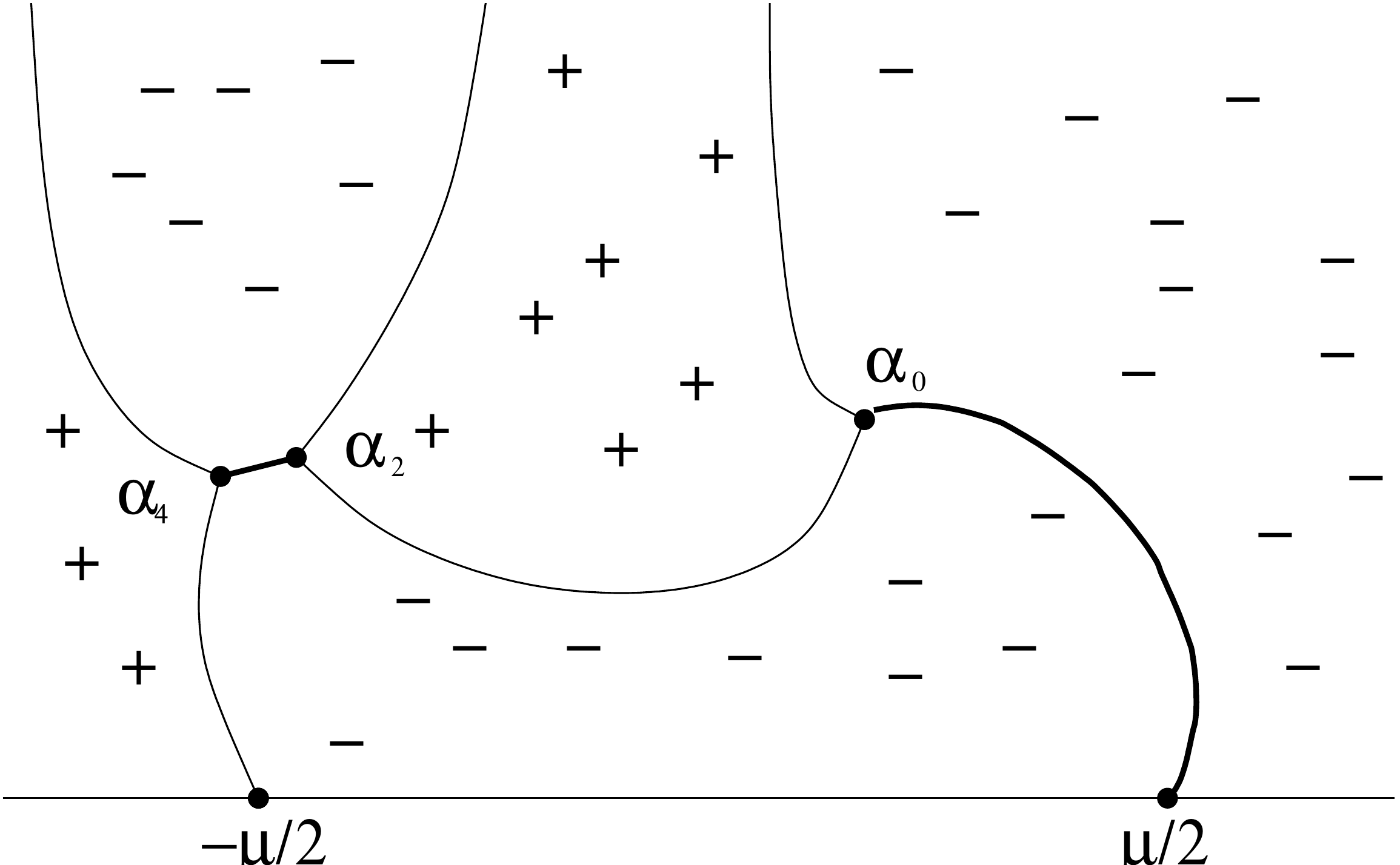}
\caption{Left: Zero level curves of $\Im h^0$, pre-break. Center: Zero level curves of $\Im h^0$; this is a breaking point
in the transition from genus zero to two;
Notice the appearance of the double point. 
Right: Zero level curves of $\Im h^2$, post-break}
\label{Break-sech}
\end{center}
\end{figure}

\bxs
1) As an example, consider the transition from genus zero to genus two for the ``sech'' potential (initial data $\cosh^{-1-\frac{2i}\varepsilon}$), represented by $f_0$
given by \eqref{f_0_sech}, see Figure \ref{Break-sech}. This transition was studied in \cite{TVZ1}. In this case, the double point $z_0$ is in the upper halfplane, 
(Figure \ref{Break-sech}, center), so the genus changes by two. The change in the  topology of zero level curves of $\Im h$,
associated with the break, is shown on Figure \ref{Break-sech}. Note that $h^0(z;x_b,t_b)\equiv h^2(z;x_b,t_b)$.

2) In the case of the box potential, the double point $z_0\in\R$ and, therefore, the genus changes from zero to one (\cite{KenJenk}). 
\exs

We  are now interested in the properties of the \blue{characteristic} velocities  at a double point $z_0$. Direct calculations  (\cite{TVZ1}) show that
 the Jacobian 
\be\label{Jac}
\frac{(\part h^N_z, \part \Im h^N)}{(\part z_0, \part t)}= |h^N_{zz}(z_0)|^2 \Im h^N_t(z_0).
\ee
If $(x_b,t_b)$ is a the regular breaking point, then $h^N_{zz}(z_0)\neq 0$. 
Let $z_0\not\in\R$. In this case 
it was  proven in \cite{TV2} that 
$|\Im h^N_t(z_0)|+|\Im h^N_x(z_0)| \neq 0$. Thus, either the Jacobian $\frac{(\part h^N_z, \part \Im h^N)}{(\part z_0, \part t)}$ or
$\frac{(\part h^N_z, \part \Im h^N)}{(\part z_0, \part x)}$ is nonzero at $z_0$. Without any loss of generality, we can assume
that the Jacobian \eqref{Jac} is nonzero at $z_0$. Then, there exists a unique breaking curve 
$t=t(x)$ passing through $(x_b,t_b)$ and a unique curve $z_0(x)$ passing through $z_0$ at $x=x_b$.
Then, differentiating  \eqref{br-cond} along $t(x),z_0(x)$, we obtain
\be\label{br-cond-a}
\Im \le(h^N_x+h^N_t\frac{dt}{dx}\ri)=0,~~~\qquad ~~h^N_{zz}\frac {dz_0}{dx}+h^N_{xz}+h^N_{tz}\frac{dt}{dx}=0.
\ee
Thus, the slope $\frac{dt}{dx}$ of the breaking curve at $x_b$ and the velocity 
\be\label{V_doub}
V=\frac{h^N_{tz}}{h^N_{xz}}=\frac{dq^N}{dp^N}
\ee
of the double point $z_0$ are
are given by
\be\label{comp-double}
\frac{dt}{dx}=-\frac{\Im h^N_x}{\Im h^N_t},~~~\qquad ~~~V=-\frac{\frac{h^N_{zz}}{h^N_{xz}}\cdot\frac {dz_0}{dx}+1}{dt/dx}
\qquad {\rm or}\qquad
\frac{dt}{dx}=-\frac{\Im p^N}{\Im q^N},~~~\qquad ~~~V=-\frac{\frac{h^N_{zz}dz}{dp^N}\cdot\frac {dz_0}{dx}+1}{dt/dx}.
\ee

In the case $z_0\in\R$, according to \eqref{br-cond}, $h^N(z)$ does not have a jump at $z_0\in \R$. Thus, it is natural to consider the case when
$h^N(z)$ is analytic in a neighborhood of $z_0$. Then, $\Im h^N\equiv 0$ on $\R$ near $z_0$, and the second condition in \eqref{br-cond}
becomes trivial. But then, if zero level curves of $\Im h^N$ are pinching $\R$ at $z_0$, we have six level curves of $\Im h^N=0$ emanating from $z_0$.
Thus, we obtain new breaking curve conditions
\be\label{br-cond-real}
h^N_z(z_0)=0~~\qquad ~~{\rm and}~~\qquad ~h^N_{zz}(z_0)=0,
\ee
which imply
\be\label{real-double}
V=-\frac{dx}{dt}, ~~~~~~\qquad \frac{dt}{dx}=-\frac{h^N_{zzz}\frac {dz_0}{dx}+h^N_{zzx}}{h^N_{zzt}}.
\ee

Thus, {\em a real double point has a real velocity}.

\bxs
 1) In the transition from genus zero (with a movable branchpoint $\a=a+ib$) to a higher genus,
 the velocity $V$ of the double point $z_0$, according to \eqref{dhdx,dhdt},  is given by $V=2z_0+\frac{b^2}{z_0-a}$,
  Using the large $t$ asymptotics of $z_0$ and of the breaking curve from \cite{TVZ2},
 we can show that in the case ``sech'' potential, of $V\not\in\R$.
 
 2) In the case of ``box'' potential, see Examples  \ref{ex-box_sech}, the double point $z_0\in \R$. The first equation 
in \eqref{br-cond-real} becomes 
\be\label{box-br-01}
4tz_0^2+2(x-L)z_0+2tq^2=0,
\ee
whereas the second equation 
in \eqref{br-cond-real} means that the discriminant of \eqref{br-cond-real} is zero.
Thus we obtain the breaking curve (see \cite{KenJenk}, \cite{box_rogue})
\be\label{br-cur01}
\frac{L-x}{t}=2\sqrt 2 q  \qquad {\rm and ~ the ~double ~point} \qquad z_0=\frac{q}{\sqrt 2}.
\ee
Notice that the double point is stationary. Further direct calculations yield $h^0_{zt}=-\frac{4z^2}{\nu(z)}$, 
$h^0_{zx}=-\frac{2z}{\nu(z)}$, so that, according to \eqref{V_doub}, $V=\sqrt 2 q $.
\exs

We now want to prove that if $\a_j$ approaches a double point $z_0$ then $V_j$
approaches the velocity $V$ of $z_0$.
 Indeed,   meromorphic differentials $dp,dq$ on the Riemann surface $\Rscr$ of genus $m\in\N$ with poles 
 at $\infty^\pm$ given by \eqref{dpq} can be
written as 
\be\label{dpdq-poly}
dp=\frac {P(z)}{R(z)}dz,~~~~dq=\frac {Q(z)}{R(z)}dz
\ee
where $P=P(z;\vec\a),~Q=Q(z;\vec\a)$ are   polynomials of degrees $m+1$ and $m+2$ respectively, where $\vec\a\in\C^{2m+2}$ 
denotes the set of $2m+2$ distinct
branchpoints of $\Rscr$. Let $\vec\a\ra\vec\b$, where $\vec \b\in\C^{2m+2}$ denotes the set $\vec\a$ after several pairs of branchpoints collided into the corresponding double points (more complicated clustering is also allowed) forming a singular Riemann surface, whose desingularization we denote by $\Rscr_0$. 
It was proven in \cite{BT3} that  Boutroux  deformations of  
$dp,dq$
are continuous in  $\vec\a\in\C^{2N+2}$ 
and 
\be\label{lim-poly}
\lim_{\vec\a\ra\vec\b}P(z;\vec\a)=\prod_j(z-z_j)P_0(z;\vec\b),     \qquad   \lim_{\vec\a\ra\vec\b}Q(z;\vec\a)=\prod_j(z-z_j)Q_0(z;\vec\b),
\ee
where $dp_0=\frac{P_0(z;\vec\b)}{R_0(z)}dz$,  $dq_0=\frac{Q_0(z;\vec\b)}{R_0(z)}dz$ are continuous limits of $dp,dq$ on $\Rscr_0$, which is 
the Riemann surface for 
 the radical $R_0$, 
$P_0,Q_0$ are the corresponding
polynomials and the product goes over all the double points  $z_j$. Then, according to \eqref{dpdq-poly}, 
\eqref{lim-poly},
\be\label{cont-V}
\lim_{\a_j\ra z_0}V_j=\le.\lim_{\a_j\ra z_0}\frac{dq}{dp}\ri|_{\a_j}=\lim_{\a_j\ra z_0}\frac{Q(\a_j;\vec\a)}{P(\a_j;\vec\a)}=\frac{P_0(z_0;\vec\b)}{Q_0(z_0;\vec\b)}
=\le.\frac{dq_0}{dp_0}\ri|_{z_0},
\ee
the latter being the velocity of the double point $z_0$.

\bigskip

{\it Acknowledgements} The work was supported  in part by the Banff International Research Station,
University of British Columbia, Vancouver,
Canada. The work of the first author was supported  in part by London Mathematical Society.
The work of the second author was supported  in part by
the Department of Mathematics, University of Central Florida, Orlando, FL 32816-1364.

\end{document}